\documentclass[prl,twocolumn,aps]{revtex4}
\usepackage{epsfig}
\begin{document}

\title{\bf Particle emission in hydrodynamics: a problem needing a solution}
\author{F. Grassi}
\address{Instituto de F\'{\i}sica, Universidade de S\~ao Paulo, 
C. P. 66318, 05315-970 S\~ao Paulo-SP, Brazil}

\begin{abstract}
A survey of various mechanisms for particle emission in hydrodynamics is 
presented.
First, in the case of sudden freeze out,
 the problem of negative contributions in the Cooper-Frye formula  
and ways out are presented.
Then the separate chemical and thermal freeze out scenario is described  and
the necessity of
its inclusion in a hydrodynamical code is discussed.
Finally, 
we show how to formulate continuous particle emission in hydrodynamics 
and
discuss extensively its consistency with data. We point out in various cases
 that the interpretation of data is quite influenced by the choice of the particle emission mechanism.

\end{abstract}
\maketitle

\section{1. Introduction}
Historically, the hydrodynamical model was suggested in 1953 by Landau
\cite{landau}
 as a way to improve Fermi statistical model
\cite{fermi}. For decades, hydrodynamics was used to describe collisions involving elementary particles and nuclei. But it really got wider acceptation with the advent of relativistic (truly) heavy ion collisions, due to the large number of particles created and its success in reproducing data.
Brazil has a good tradition with hydrodynamics. 
Many aspects of it have been treated by various persons. 
For illustration, the following papers can be quoted.
Initial conditions were studied in \cite{ha97a,ag02}. Solutions of the hydrodynamical equations 
using symmetries \cite{ha85,cs03} or numerical
\cite{el99,ag01} were investigated. The equation of dense matter was
derived in  \cite{me93,ag03}.
Comparison with data was performed in \cite{ha88,pa98,pa02,gr00,gr98,gr99a,
gr99b}. 
The emission mechanism was considered in \cite{ha91,na92,gr95,ma99,ar01,yogiro}.
In this paper, I concentrate on the problem of particle emission in 
hydrodynamics. 	In the Fermi description, energy is stored in a small volume, 
particles are produced according to the laws of
statistical equilibrium at the instant of equilibrium  and they immediately stop interacting, i.e. they freeze out. Landau took up these ideas: 
energy is stored in a small volume, 
particles are produced according to the laws of
statistical equilibrium at the instant of equilibrium, expansion occurs 
(modifying particle numbers in agreement with the laws of conservation)
and stops when the mean free path becomes of order the linear dimension of the system, which  led to a decoupling temperature of order the pion 
mass for a certain energy and slowly decreasing with increasing energy.
In today's hydrodynamical description, two Lorentz contracted nuclei collide.
 Complex processes take place in the initial stage leading to a state of 
thermalized hot dense matter at some  proper time $\tau_0$. This matter evolves according to the laws of hydrodynamics. 
As the expansion proceeds, the fluid becomes cooler and more diluted
 until interactions stop and particles free-stream towards the detectors.
In the following, I review various possible descriptions for this last stage of the hydrodynamical description. 
The usual mechanism for particle emission in hydrodynamics
is sudden freeze out so I will use it as a  point of comparison.  I will start in section 2,
 reminding what it is, some of its problems and
ways outs.
There is another particle emission scenario which is a small extension of this idea of sudden freeze out: the separate chemical and thermal freeze out
scenario. It has become used a lot e.g. to analyse data. So I will discuss 
in section 3
what it is, its alternatives and how to incorporate it in hydrodynamics.
Continuous emission is a  
mechanism for particle emission that we proposed some years ago.
As the very name suggests, it is not ``sudden'' like the usual
freeze out mechanism. I will explain what it is precisely in
section 4 and how  it describes data compared to freeze out.
Finally I will conclude in section 5.

\section{2. Sudden freeze out}

\subsection{2.1 The traditional approach and its problems}

Traditionally in hydrodynamics, the following simple picture is used. Matter expands until a certain dilution criterion is satisfied. Often the criterion used is that a certain temperature has been reached, typically around 140 MeV in the spirit of Landau's case. In some more modern version such as \cite{hi}, 
a certain freeze out density must be reached. There also exist 
attempts \cite{he87,le88,hu98,leh90,na92}
to incorporate more physical informations about the freeze out, for example 
type i particles stop interacting when their average time between interactions $\tau^i_{scatt}$ becomes greater than the fluid expansion time and 
average time to reach the border.
When the freeze out criterion is reached, it is assumed that all particles stop interacting suddenly (this is called ``freeze out'') and fly freely towards the detectors. As a consequence, observables only reflect the conditions 
(temperature, chemical potential, fluid velocity) met by matter late in its evolution.

In the sudden freeze out model, to actually compute particle spectra and
get predictions for the observables, 
the Cooper-Frye formula (\ref{eq:CF}) \cite{CFart} may be used.

\begin{equation}
\frac{E d^3N}{dp^3}= 
\int_{T_{f.out}} d\sigma_{\mu} p^{\mu} f(x,p).
\label{eq:CF}
\end{equation} 
 $d\sigma_{\mu}$ is the normal vector to this surface,
$p^\mu$ the particle momentum
and $f$ its  distribution function. Usually one assumes a Bose-Einstein or Fermi-Dirac distribution for this $f$.

This sudden freeze out approach is often used however it is known to have some bad features. I will mention two. First when using the Cooper-Frye formula,
we sometime meet negative terms ($d\sigma_{\mu} p^\mu \leq 0$) corresponding to particles re-entering the fluid. However since they presumably
had stopped interacting (being in the frozen out region),
they should not re-enter the fluid and start interacting again.
\begin{figure}[htbp]
\begin{center}
\epsfig{file=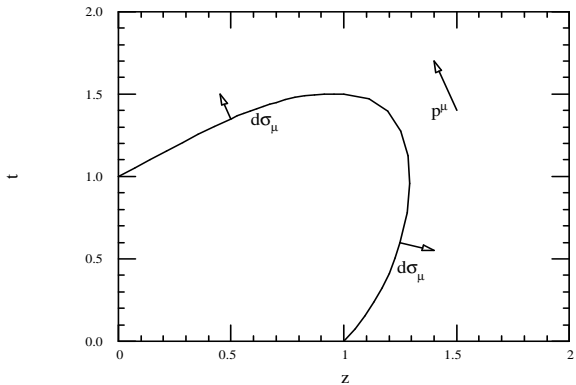,height=8.5cm,angle=-90} 
\end{center}
\caption[CF]{In Cooper-Frye formula, the expression
$p^{\mu}d\sigma_{\mu}$ may be negative.
\label{fig:cffig}}
\end{figure}

\begin{figure}[htbp]
\begin{center}
\epsfig{file=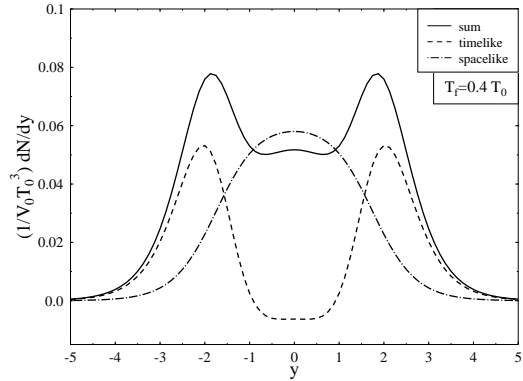,height=7.cm,angle=-90}
\end{center}
\caption[Negative contributions]{ 
Rapidity distribution of particles freezing out
on the
$T_{f.out}=0.4 T_0$ isotherm in the Landau model. The solid line corresponds
to all the  contributions in  formula 1, the dashed-dotted line
represents  contributions from space-like parts of the isotherm and the dashed line
 contributions from time-like  parts  \cite{be96,ri98}.
In this last case, note the negative contributions at y=0.
\label{fig:rischke}}
\end{figure}

So usually one removes these negative terms from the calculations as being unphysical.
However by doing this, one removes 
baryon number, energy and momentum from the calculation and violates
 conservation laws. It is not a negligible problem, as shown in the figure~\ref{fig:rischke}.
In the code SPHERIO~\cite{ag01},
it can be a 20\% overestimate of particle number. 
 There are some ways to avoid these violations  but none is completely satisfying~\cite{ma99}.


The second problem is the following: do particles really {\em suddenly} stop 
interacting when they reach a certain hypersurface? Intuitively no, 
this must happen over a mean free path. This is corroborated by results from
simulations of microscopical models\cite{Bra,Sor,Bas}: 
the shape of the region where particles last interacted is generally not a sharp surface as assumed for sudden 
freeze outs. Some exceptions might be heavy particles in heavy systems or the
phase transition hypersurface.

We postpone the discussion of the second problem to section 4 and turn to the first problem.

\subsection{2.2 Improved freeze out}

In this section, we adopt the sudden freeze out picture and seek ways to incorporate conservation laws~\cite{ma99}.

\begin{figure}[htbp]
\begin{center}
\epsfig{file=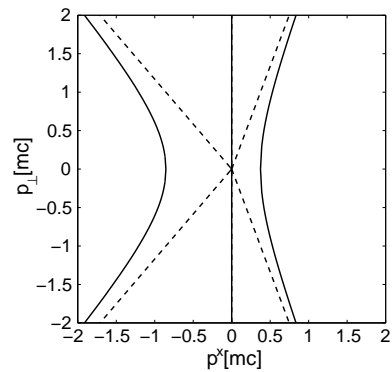,height=5.cm}
\caption{ The solution to the equation  $d\sigma_{\mu} p^{\mu}=0$ 
for
$v$
= 0,65, 0 and -0,35 is given by the left  hyperbola, the vertical axis
and right hyperbola respectivaly. 
 The permitted region is 
localized to the right of the curve in each of these three cases.
The dashed lines  correspond to the  case of
massless  particles.}
\label{fig:selec}
\end{center}
\end{figure}

We suppose that prior to crossing the freeze out surface $\sigma$, particles have a thermalized distribution function and we know the baryonic current and energy momentum tensor, $n_0^{\mu}$ and $T_0^{\mu \nu}$. We suppose also that after crossing the surface, the distribution function is
\begin{equation}
f^*_{FO}(x,p;d\sigma^\mu)  
= 
f_{FO}(x,p) \Theta(p^\mu\ d\sigma_\mu) 
\label{eq:cut}
\end{equation}
The $\Theta$ function selects among particles which are emitted only 
only those with $d\sigma_{\mu} p^{\mu}>0$.
This equation is solved in the rest frame of the gas doing
freeze out (RFG) 
in figure~\ref{fig:selec} . We see that according to the value of
 $v={d\sigma_0/
d\sigma_x}_{RFG}$, 
a region more or less big in the
 $p_{\perp}-p^x$ space can be excluded. 

We do not know what the shape of
$f_{FO}(x,p)$ is. To simplify, we first suppose that
\begin{equation}
f_{FO}(x,p)  
= \frac{1}{(2 \pi)^3} exp \left( \frac{-p_\mu u^\mu+\mu}{T} \right)
\end{equation}
with $u^\mu=\gamma(1,v,0,0)$ and $\mu$ is the baryonic potential.\\ 
This does not mean that $f_{FO}$ is thermalized but simply that
we choose a parametrization of the thermalized type.
This parametrization is arbitrary, we discuss later how to improve our ansatz.
For the moment we use it to illustrate how to proceed in order not to violate conservation laws when using the Cooper-Frye formula.

It is possible to find expressions for the baryonic current, energy momentum
tensor and entropy current corresponding to \ref{eq:cut}
in terms of Bessel-like functions  and for massless particules, even analytical
expressions as function of $v$, $T$ e $\mu$ \cite{ma99}.

\begin{figure}
\begin{center}
\epsfig{file=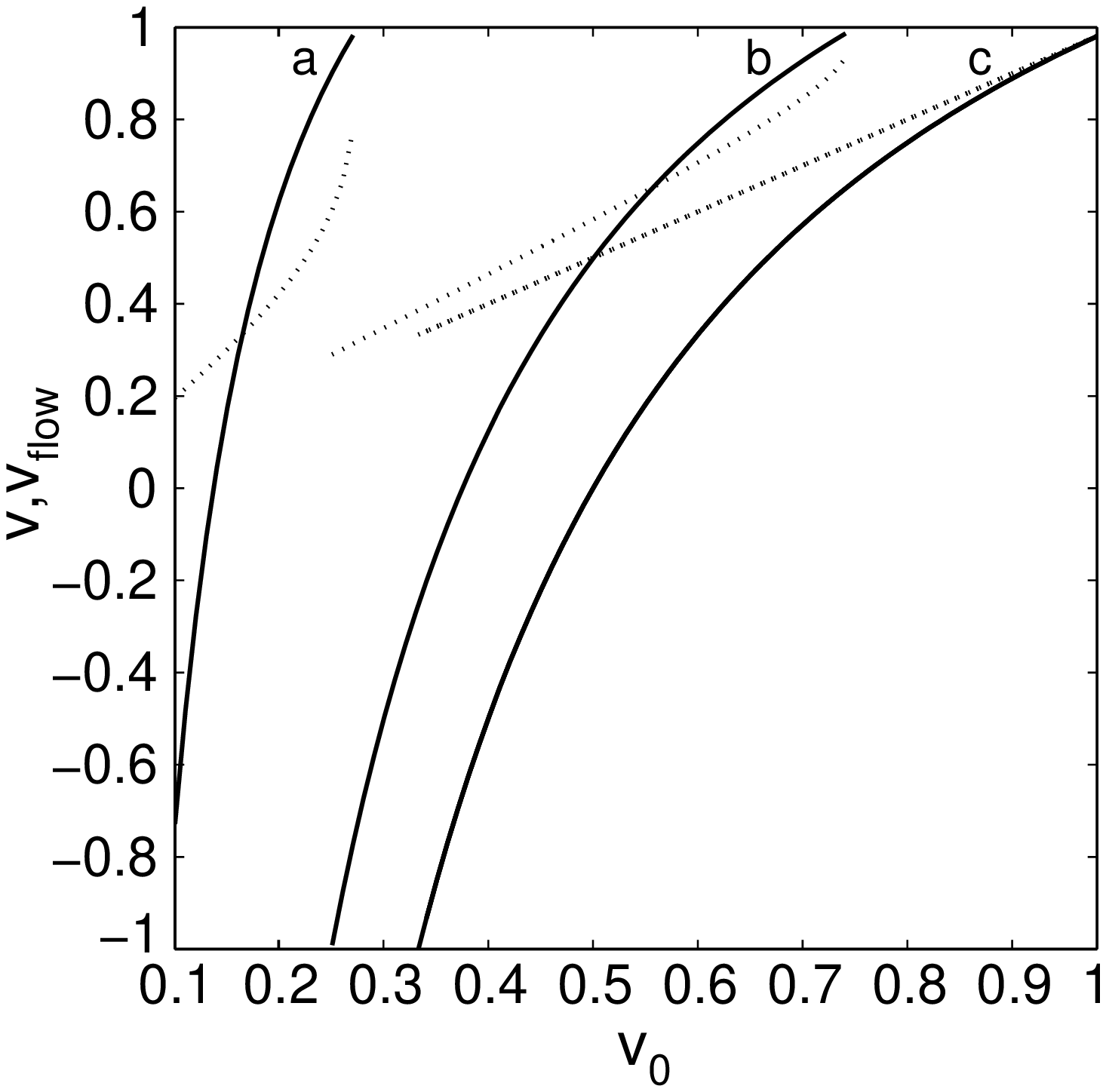,height=5.5cm}\hfill
\epsfig{file=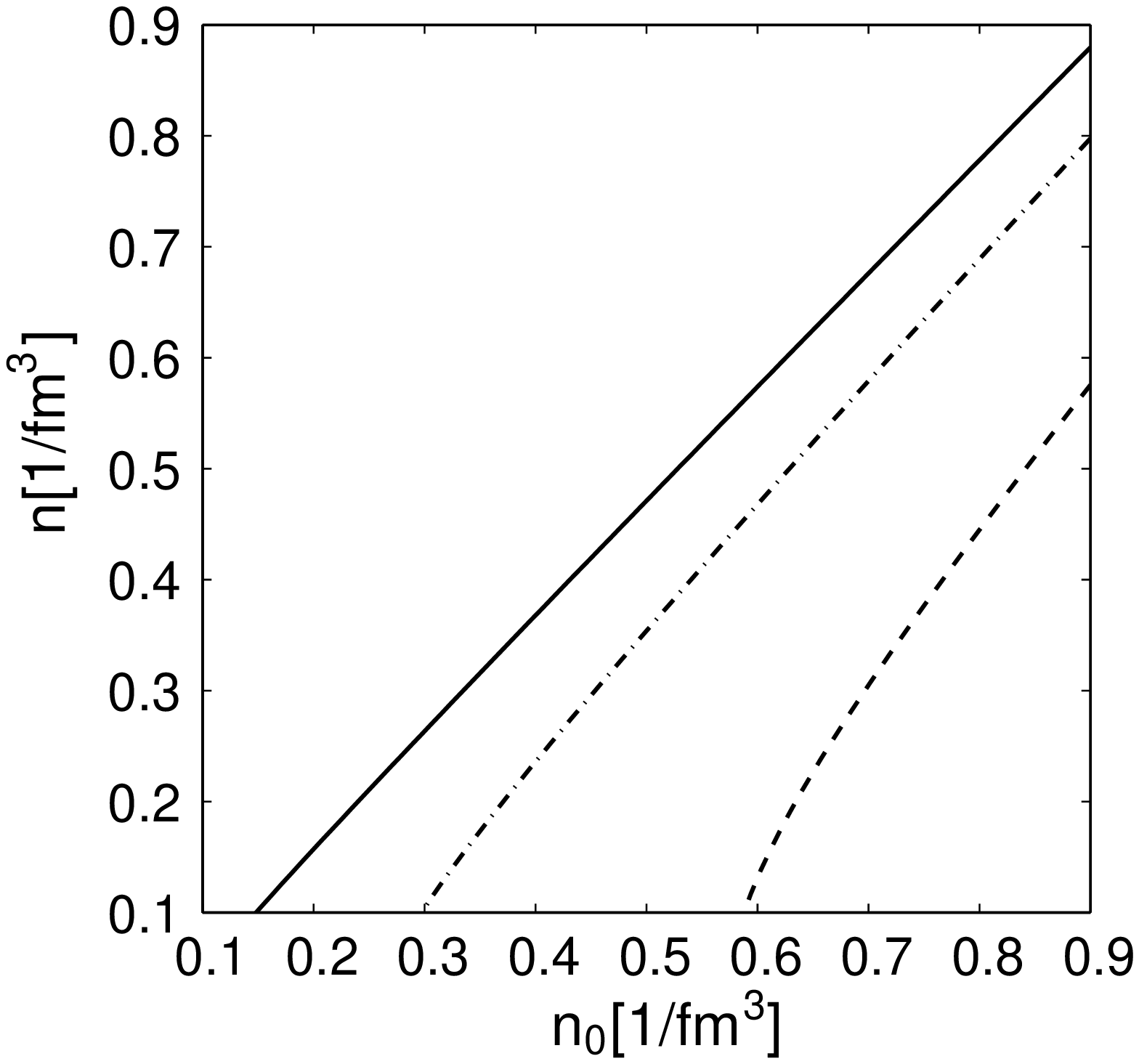,height=5.5cm}\hfill
\epsfig{file=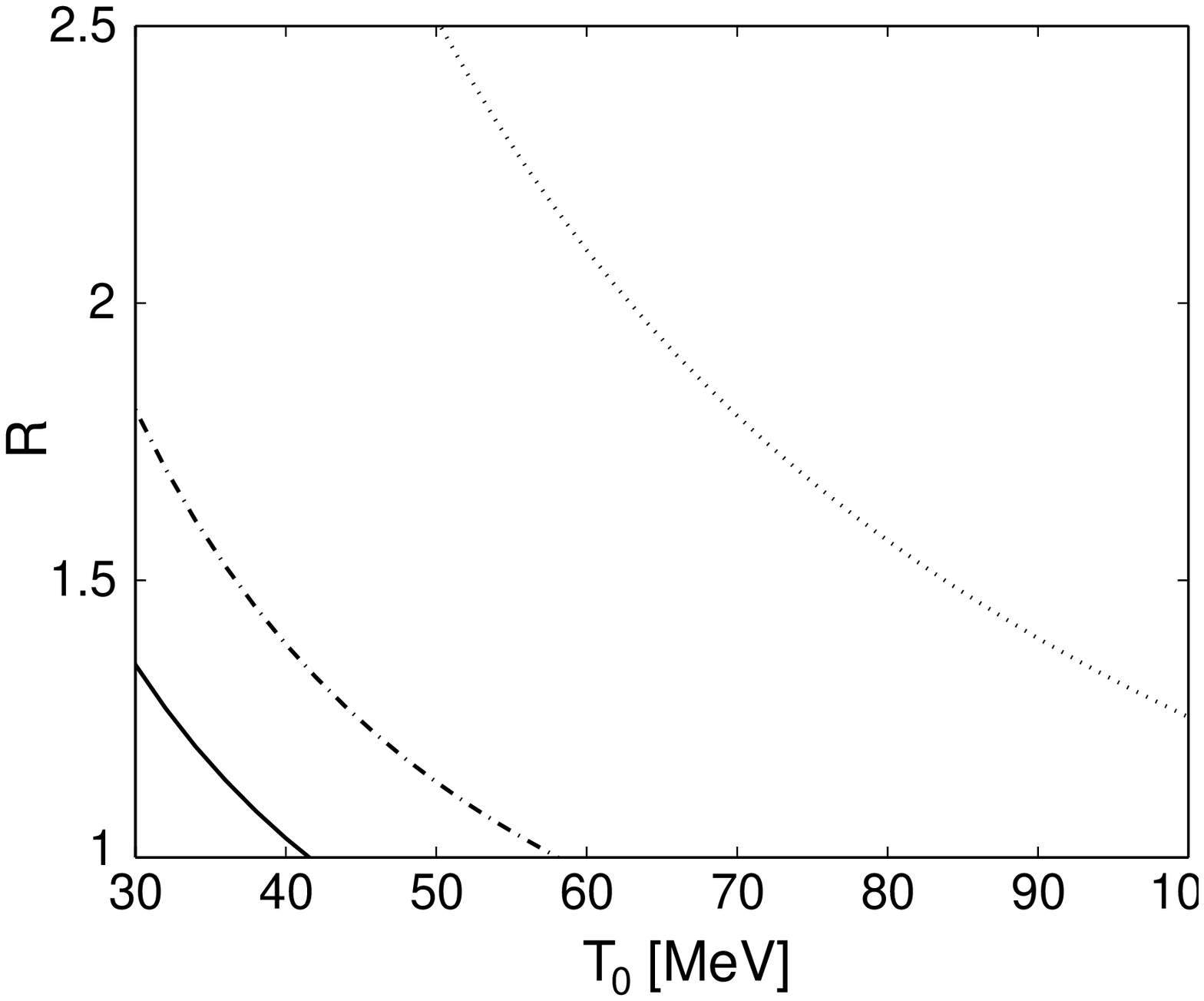,height=5.5cm}
\end{center}
\caption[solution]{ Solution of conservation laws in the case of a plasma.\\
Top: $v$ as function of 
$v_0$ (solid line) for 
a) $n_0=1.2\,fm^{-3}, T_0=60\,MeV, \Lambda_B \equiv B^{1/4}=225\,MeV$,
b) $n_0=0.1\,fm^{-3}, T_0=60\,MeV, \Lambda_B =80\,MeV$,
c) $n_0=1.2\,fm^{-3}, T_0=60\,MeV, \Lambda_B =0\,MeV$.
(Dashed lines: velocity of pos-freeze out baryonic flow). \\
Middle: baryonic density
$n$ as function of $n_0$ for $v_0=0.5$, $T_0=50\,MeV$,
a) $\Lambda_B =80\,MeV$ (continuous line), 
b) $\Lambda_B =120\,MeV$ (dashed-dotted line), 
c) $\Lambda_B =160\,MeV$ (dashed line).\\
Bottom : $R$, ratio of entropy currents for   post and
pre-freeze matter as  function of  $T_0$ for
a) $n_0=0.1\,fm^{-3},v_0=0.5\,MeV, \Lambda_B =80\,MeV$ (solid line), 
b) $n_0=0.5\,fm^{-3},v_0=0.5\,MeV, \Lambda_B =80\,MeV$ (dashed line),  
c)  $n_0=1.2\,fm^{-3},v_0=0.5\,MeV, \Lambda_B =225\,MeV$ (dotted line).}
\label{fig:conserv}
\end{figure}

To determine the parameters $v$,T and $\mu$ for matter on the post-freeze out 
side of  $\sigma$,
we need to solve the conservation equations\\
\centerline{$[N^\mu d\sigma_\mu]=0\,\,\,\,\,\,\,\,[T^{0\mu}d\sigma_\mu]=0\,\,
\,\,\,\,\,\,[T^{x\mu}d\sigma_\mu]=0$,}
as function of quantities  for matter on the pre-freeze out side,
 $v_0$, $T_0$ e $\mu_0$.
This being done, we still need to check that\\
\centerline{$[S^\mu d\sigma_\mu]\geq0$ or $R=\frac{S^\mu d\sigma_\mu}
{S_0^\mu d\sigma_\mu} \geq 1$}
i.e. entropy can only increase when crossing  $\sigma$.
Generally, these equations need to be solved numerically but for massless particles, they have an analytical solution  \cite{ma99}.
For illustration, we show this solution for the case of a plasma
with an MIT bag equation of state in figure~\ref{fig:conserv}.
An interesting result can be seen on the top figure. Normally
when  using a  Cooper-Frye formula, the velocity of matter 
pre and post freeze out is assumed to be the same.
 However in the  figure, one sees that when imposing conservation laws,
 matter may be  acelerated in a  subtantial way. For example in case a), 
$v_0=0.2$ implies $v_{flow}=0.4$ and $v=0.6$. 
In term of effective temperature, there was an increase of
 60\%.

This example illustrates the  importance of taking into account
conservation laws when crossing $\sigma$. However, the choice of
 $f_{FO}$ as being parametrized
in the same way as a thermalized distribution is arbitrary as we mentioned already.
So we now study a more physical way of computing this function.

Consider an infinite tube with the $x<0$ part filled with  matter and the
 $x>0$ part empty. At t=0, we remove the  partition at  $x=0$ and  matter
 expands in  vacuum.
Suppose we remove the particles on the right hand side and put them back on the
left hand side continuously so as to get a stationary flow, with a rarefaction wave propagating to the left of the matter.
 
In the spirit of the continuous emission model presented below, 
the distribution function of matter has two
 components,
$f_{free}$ and $f_{int}$. Suppose that $f_{free}(x=0,p)=0$  and
$f_{int}(x=0,p)=f_{therm}((x=0,p)$.
A simple  model  for the fluid evolution is 
\begin{eqnarray}
\partial_x f_{int}(x,\vec{p})   dx &=& - \Theta(p^\mu d\hat{\sigma}_\mu) 
                                   \frac{\cos \theta_{\vec{p}} }{\lambda}
           f_{int}(x,\vec{p})   dx,\nonumber
 \   \\ 
\partial_x f_{free}(x,\vec{p})  dx &=& + \Theta(p^\mu d\hat{\sigma}_\mu) 
                                   \frac{\cos \theta_{\vec{p}} }{\lambda}
           f_{int}(x,\vec{p})   dx.
\end{eqnarray}
where $\cos \theta_{\vec{P}} = \frac{p_{x}}{p}$ in the rest frame of the rarefaction wave. A solution for these equations is
\begin{equation}
f_{int}(x,\vec{p}) =  f_{therm}(x=0,\vec{p}) 
\exp \left[ - \Theta(p^\mu d\hat{\sigma}_\mu) 
              \frac{\cos \theta_{\vec{p}} }{\lambda}  x \right].
\end{equation}
and
\begin{eqnarray}
f_{free}(x,\vec{p}) & = & f_{therm}(x=0,\vec{p}) \times \nonumber \\
& & \left\{1-\exp\left[-\Theta(p^\mu d\hat{\sigma}_\mu)
\frac{\cos\theta_{\vec{p}}}{\lambda}x\right]\right\} \nonumber \\
& = & f_{therm}(x=0,\vec{p})  - f_{int}(x,\vec{p}) 
\end{eqnarray}
We see that  $f_{free}$  tends to the cut  thermalized distribution we saw above when 
  $x \longrightarrow \infty$.
In this model,  the particle density does not change with $x$ but 
particles with  {$p^\mu d\hat{\sigma}_\mu  > 0$
 pass gradually from  $f_{int}$ to $f_{free}$. 
 
To improve this model, we consider
\begin{eqnarray}
\partial_x f_{int}(x,\vec{p}) dx & = & - \Theta(p^\mu d\hat{\sigma}_\mu) 
                                   \frac{\cos \theta_{\vec{p}} }{\lambda}
           f_{int}(x,\vec{p})   dx \nonumber \\
                   &  &+\left[ f_{eq}(x,\vec{p}) -  f_{int}(x,\vec{p})\right]
           \frac{1}{\lambda'} dx, 
\end{eqnarray}

\begin{equation}
\partial_x f_{free}(x,\vec{p}) dx = + \Theta(p^\mu d\hat{\sigma}_\mu) 
                                   \frac{\cos \theta_{\vec{p}} }{\lambda}
           f_{int}(x,\vec{p})  dx.
\end{equation}
The additional term in
$f_{int}$ includes the tendency for this function to tend towards 
an equilibrium function due to collisions
with a relaxation distance  $\lambda'$.
Due to the loss of energy,
 momentum and  particle number, $f_{eq}$ is not the initial thermalized 
function but its parameters
$n_{eq}(x)$, 
$T_{eq}(x)$ and $u^\mu_{eq}(x)$ can be  determined using conservation laws.
In the case of immediate re-thermalization ($f_{int}=f_{eq}$), for
a gas of massless particles with zero net baryon number, the solution is shown in
 figure~\ref{fig:dist}.
\begin{figure}[htbp]
\begin{center}
\epsfig{file=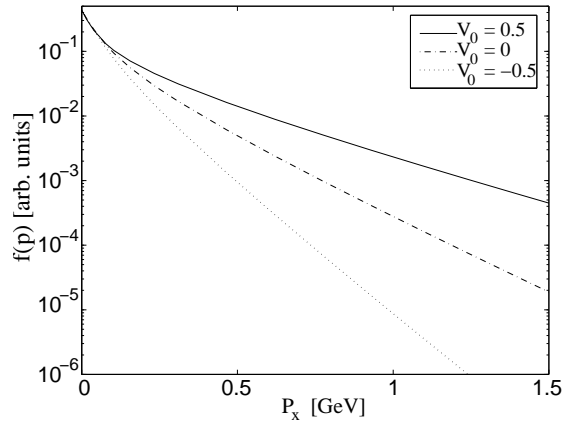,height=6.cm}
\caption{ $f_{free}(p_x)$ ( equivalent to the
$f^*_{FO}$ in the previous section), computed at
$p_y=0, x=100\lambda, T_0=130\,MeV$.}
\label{fig:dist}
\end{center}
\end{figure}
One sees that the
 solution $f_{free}$  is not a thermalized type cut function.

Even more importantly, this distribution exibits a  curvature which reminds
the data on  $p_{\perp}$ distribution for  pions. 
Other explaination for this  curvature are transverse
 expansion or  resonance decays. On the basis of our work, it is difficult
to trust totally  analyses
which extracts thermal freeze out temperatures and fluid  velocities
 using only  transverse expansion and resonance decays. 

More details and
improvement on how to compute the post freeze out distribution function were 
presented in various papers \cite{an99,ma99b,ma03,cs04,ta04}.

Finally, it is interesting to note that models that combine hydrodynamics with 
a cascade code  also suffer from the problem of non-conservation of energy-momentum and charge \cite{tea01} or inconsistency
\cite{du00}, as discussed in \cite{bu03}.

 
\vspace*{0.5cm}

\section{3. Separate freeze outs}

In this section, we suppose that the sudden freeze out picture can be used and see how well it describes data.

\subsection{3.1 Chemical freeze out}

Strangeness production plays a special part in ultrarelativistic
nuclear collisions since its increase might be evidence for the creation
of quark gluon plasma. Many experiments therefore collect information on strangeness production.

We can consider for example the results obtained by the CERN collaborations
WA85 (collision S+W),  WA 94 (collision S+S) and WA97, later on NA57 (collision Pb+Pb).
One can combine various ratios to obtain a window for the freeze out conditions
compatible with all these data. The basic idea is simple, for example:
\begin{equation}
\overline{\Lambda}/\Lambda \sim 
\exp \frac{2 (\mu_S-\mu_B)}{T}_{|f.out}=exp.\,value.
\end{equation}
(neglecting decays.)\\

In principle, this equation depends on three variables. However, supposing that strangeness is locally conserved, this leads to a relation $\mu_S(\mu_b,T)$, then given a minimum and a maximum values, the above equation gives a relation 
$T_{f.out}(\mu_{B\,f.out})$. We show typical results in figure~\ref{fig:redlich}.
\begin{figure}[htbp]
\begin{center}
\hspace*{-2.cm}
\epsfig{file=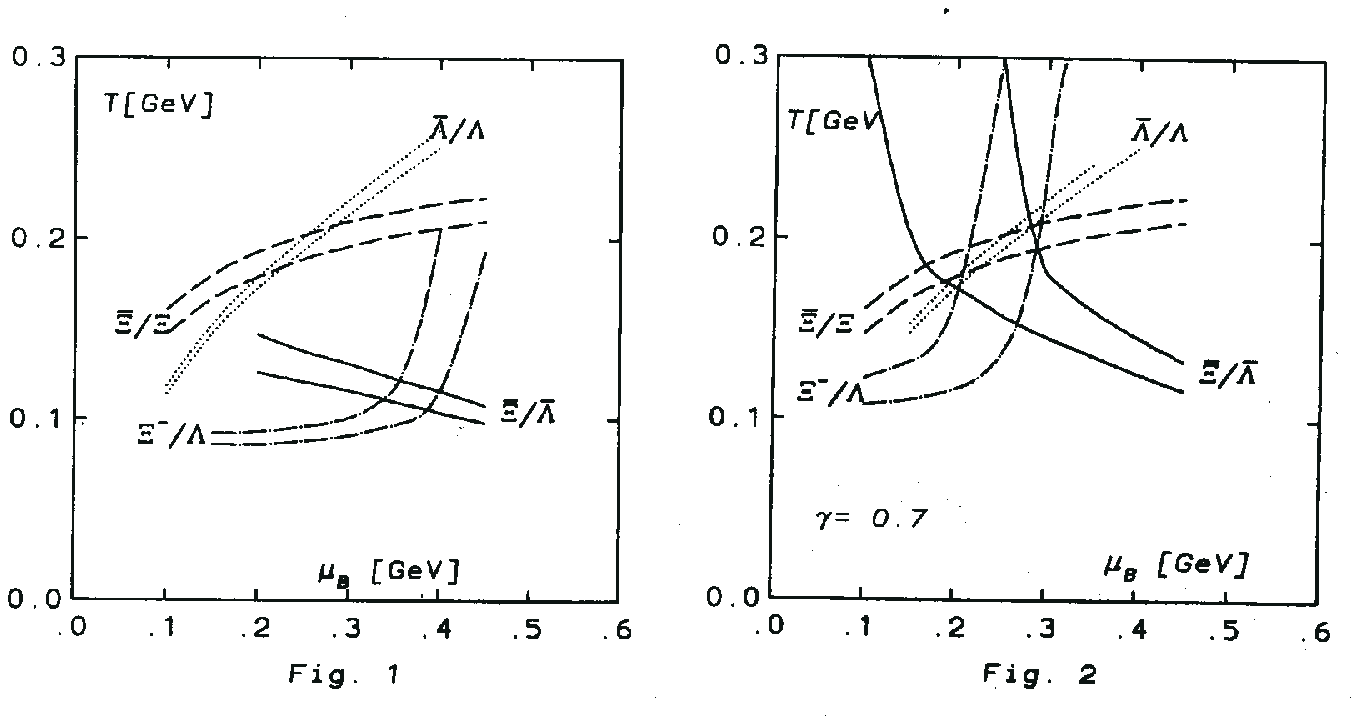,height=6.cm}
\end{center}
\caption{ Search of a window \cite{re94} of values of $T_{f.out}$ e
  $\mu_{b,\,f.out}$ reproducing WA85 data. This window only exists for
$\gamma < 1$ (for this reference).
\label{fig:redlich}}
\end{figure}
The parameter $\gamma$ in this figure is basically a 
phenomenological factor, which 
indicates how far from chemical equilibrium we are, it is introduced
in front of the factors
 $e^{\pm \mu_s/T}$ where $\mu_s$ is the s quark chemical potential, 
 $\mu_s=\mu_B/3-\mu_S$.
(There exists a study by C.Slotta et al. \cite{cl95}
 motivating this way of  including $\gamma$.)
It can be seen that if
 $\gamma=0,7$ it is possible to
reproduce the experimental ratios
  $\overline{\Lambda}/\Lambda$,
 $\overline{\Xi}/{\Xi}$,
$\Xi/\Lambda$, $\overline{\Xi}/\overline{\Lambda}$ choosing 
$T_{f.out}$ and $\mu_{b,\,f.out}$ in a certain window.
This window is located around 
$T_{f.out}\sim 180-200$ MeV, $\mu_{b,\,f.out}\sim 200-300$ MeV.
One can be surprised by such high values since particle densities
are still high.
However these results are rather typical as can be checked in the table
~\ref{tab:f.chem.}.
\begin{table}[htbp]
\caption{ Summary of freeze out values from J.Sollfrank
  \cite{so97a}.}
\label{tab:f.chem.}
\begin{center}
 \begin{tabular}{lllll}\hline 
collision & T (MeV) & $\mu_b$ (MeV) & $\gamma_s$ & ref. \\ \hline
S+S       & 170     &  257          &      1     & \cite{Davidson91a} \\
       & 197$\pm$29 & 267$\pm$21 & 1.00$\pm$0.21 & \cite{so94} \\
          & 185     &  301          &      1     &  \cite{tawai96}\\
       & 192$\pm$15 & 222$\pm$10    &      1     & \cite{panagiotou96} \\
       & 182$\pm$9  & 226$\pm$13 & 0.73$\pm$0.04 & \cite{becattini97} \\
       & 202$\pm$13 & 259$\pm$15 & 0.84$\pm$0.07 & \cite{so97b} \\ \hline
S+Ag   & 191$\pm$17 & 279$\pm$33    &      1     & \cite{panagiotou96} \\
    & 180.0$\pm$3.2 & 238$\pm$12 & 0.83$\pm$0.07 &  \cite{becattini97} \\
       & 185$\pm$8  & 244$\pm$14 & 0.82$\pm$0.07 & \cite{so97b} \\ \hline
S+Pb   & 172$\pm$16 & 292$\pm$42    &      1     & \cite{andersen94} \\
S+W    & 190$\pm$10 & 240$\pm$40    &     0.7    &  \cite{re94}  \\
       & 190        & 223$\pm$19 & 0.68$\pm$0.06 &  \cite{letessier95}\\
       & 196$\pm$9  & 231$\pm$18 & 1             &  \cite{braun96}\\
S+Au(W,Pb) & 165$\pm$5 & 175$\pm$5 &      1      &  \cite{panagiotou96} \\
           & 160    &  171       &        1      &  \cite{spieles97}\\
           & 160.2$\pm$3.5 & 158$\pm$4 & 0.66$\pm$0.04 &  \cite{so97b} \\ \hline        
\end{tabular}
\end{center}
\end{table}

The
 explaination usually given nowadays is that at these temperatures, particles
 are doing a {\em chemical freeze out}, they stop having inelastic collisions and their abundances are frozen. 

\subsection{3.2 Thermal freeze out}

Transverse mass distributions obtained experimentally 
(when plotted logarithmically) exhibit large inverse inclinations.
These are called effective temperatures. 

In the  case of hydrodynamics, these effective temperatures are thought
to be due to the convolution of the fluid temperature with its transverse velocity, both at freeze out. So in particular the effective temperatures are higher than the fluid temperature.
In addition, the effective temperature should be larger the larger the 
particle mass is , since the ``kick'' received  in momentum,
$\sim m v^{fluid}_{f.out}$,
 due to transverse 
expansion,
  is larger (this argument is only valid for the non-relativistic part of the spectrum i.e. $p_{\perp}<<m$; the effective temperature does not depend on mass for the part of the spectrum where
$p_{\perp}>>m$ (but for that part of the spectra other phenomena might be important).
In general, given an experimental 
 $m_{\perp}$ spectrum, there exist many pairs 
of   $T_{f.out}$ and   $v^{fluid}_{f.out}$ which can reproduce it.

To remove this ambiguity, we can compare the $m_{\perp}$ spectra for various
types of particles (e.g. \cite{na44}), or for a given type of particle, for example pions, 
combine the fit of the spectrum with results on HBT correlations
(e.g. \cite{wi97}).
A compilation for various accelerator energies 
of values for $T_{f.out}$   obtained from particle spectra
is presented in figure \ref{fig:he97fig} (dashed line).
\begin{figure}
\begin{center}
\epsfig{file=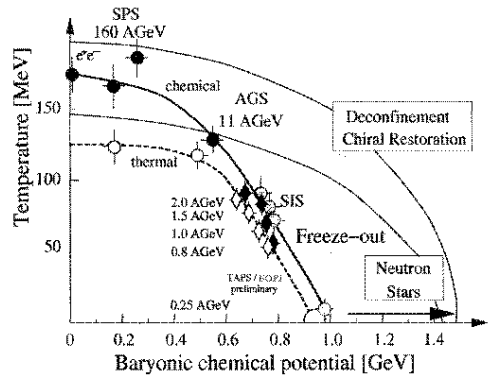,height=5.cm,angle=0}
\caption{Phase diagram with lines indicating values of freeze out parameters
at various energies
obtained from particle abundances (solid line) and particle spectra (dashed line) \cite{he97}. }
\label{fig:he97fig}
\end{center}
\end{figure}
A typical value
 at SPS is 
$T_{f.out}\sim 125$ MeV. 
One can be surprised by the fact that
these values for $T_{f.out}$ are lower than those obtained from particle ratios.
The usual explaination for this is that 125 MeV corresponds to a thermal freeze out of particles, i.e. when they stop having elastic collisions and the shape of their spectra become frozen.
This model with a chemical freeze out followed by a thermal freeze out is called separate freeze out model. 
Its parameters depend on the energy as shown in figure~\ref{fig:he97fig}; the
possible decrease of  $T_{f.out}$ with increasing energy is discussed
 in \cite{ha91,na92}.
Some possible problems of this model ($\Omega$ temperature and pion abundance are discussed later).
For the moment let us see how to incorporate this description in hydrodynamics, since so far it was based only on the analysis of two different types of data
with thermal or semi-analytical hydrodynamical models.

\subsection{3.3 Is it quantitatively  necessary to modify hydrodynamics to
incorporate separate freeze outs?}

In \cite{ar01}, we made a preliminary study of whether such a
separate freeze out model would quantitatively influence the hydrodynamical expansion of the 
fluid. For this, we used a simple hydrodynamical model, with longitudinal expansion only and longitudinal boost invariance~\cite{bj}.

For a single freeze out, the hydrodynamical equations are
\begin{eqnarray}
\frac{\partial \epsilon}{\partial t}+\frac{\epsilon+p}{t} & = & 0 \nonumber\\
\frac{\partial n_B}{\partial t} + \frac{n_B}{t} & = & 0
\label{eq:bj}
\end{eqnarray}
The last equation can be solved easily
\begin{equation}
n_B(t)=\frac{n_B(t_0) t_0}{t}.
\end{equation}

Given an equation of state,
 $p(\epsilon,n_B)$, we can get
 $\epsilon(t)$ and $n_B(t)$, solving (\ref{eq:bj}). 
From them,   $T(t)$ and 
 $\mu_B(t)$ can be extracted. 
So, if the
freeze out criterion is $T_{f.out}=constant$, one can see which are the values
for other quantities at
freeze out, for example
the  values of  $t_{f.out}$, $\mu_{B\,f.out}$, ... These values being known,
spectra can be computed.

Now let us start again with the previous model, but we suppose
that when a certain temperature $T_{ch.f.out}$ is reached
  (corresponding to a certain $t_{ch.f.out}$) some abundances are
frozen.
To fix ideas, let us suppose that  $\Lambda$ and
$\bar{\Lambda}$ are in this situation.
In this case, for
 $t \geq t_{ch.f.out}$,
in addition 
to the  hydrodinamic equations above, (~\ref{eq:bj}), 
we must introduce separate conservation laws
 for these two types of particles
\begin{eqnarray}
\frac{\partial n_{\Lambda}}{\partial t} + \frac{n_\Lambda}{t} & = & 0\\
\frac{\partial n_{\bar{\Lambda}}}{\partial t} + \frac{n_{\bar{\Lambda}}}{t} & = & 0.
\end{eqnarray}
Again it is easy to solve these equations
\begin{eqnarray}
n_{\Lambda}(t)& = & \frac{n_{\Lambda}(t_0) t_0}{t}\\
n_{\bar{\Lambda}}(t) & = & \frac{n_{\bar{\Lambda}}(t_0) t_0}{t}
\label{eq:dens}
\end{eqnarray}
For times  $t \geq t_{ch.f.out}$, 
we need to solve the  hydrodynamic equations 
 (~\ref{eq:bj}) 
with an equation of state modified to
  incorporate these conserved  abundances.

We suppose that the fluid is a gas of non-interacting resonances.
\begin{eqnarray}
n_i & = & \frac{g_i m_i^2 T}{2 \pi^2} \sum_{n=1}^{\infty} (\mp)^{n+1}
\frac{e^{n \mu_i/T}}{n} K_2(n m_i/T)\\
\epsilon_i & = & \frac{g_i m_i^2 T^2}{2 \pi^2} \sum_{n=1}^{\infty} (\mp)^{n+1}
\frac{e^{n \mu_i/T}}{n^2}[3 K_2(n m_i/T)\\
 & + & \frac{n m_i}{T} K_1(n m_i/T)]
\nonumber
\\
p_i  & = & \frac{g_i m_i^2 T^2}{2 \pi^2} \sum_{n=1}^{\infty} (\mp)^{n+1}
\frac{e^{n \mu_i/T}}{n^2} K_2(n m_i/T)
\end{eqnarray}
where $m_i$ is the particle mass, $g_i$, its degeneracy and
$\mu_i$, its chemical potential, the minus sign holds for fermions and
plus for bosons. In principle each particle
species  $i$ making early chemical freeze-out has a chemical potential
associated to it; this potential controls the conservation of the
number of particles of type $i$. For particle species not making early
chemical freeze-out, the chemical potential is of the usual type,
$\mu_i=B_i \mu_B +S_i \mu_S$, where $ \mu_B$ ($\mu_S$) ensures the
conservation of baryon number (strangeness) and $B_i$ ($S_i$) is the baryon
(strangeness)
number of particle of type $i$. 
So the modified equation of state depends not only on $T$ and $\mu_B$ but also
$\mu_{\Lambda}$, $\mu_{\bar{\Lambda}}, etc $. 
(the notation ``$etc$''  stands for all the other particles making early
chemical freeze-out). This complicates  the
 hydrodynamical problem, however we can note the following.

If $m_i-\mu_i>> T$ (the density of type $i$ particle is low)
and  $m_i >> T$, (these relations should hold for
all particles except pions and we checked them  for various times and
particle types)
the
following
approximations can be used
\begin{eqnarray}
n_i & = & \frac{g_i}{2 \pi^2} \sqrt{\frac{\pi}{2}} (m_i T)^{3/2}
e^{(\mu_i-m_i)/T} \nonumber \\
& & \times \left(1+\frac{15 T}{8 m_i }
                  + \frac{ 105 T^2}{128 m_i^2}+...
\right)
\\
\epsilon_i & = & n_i m_i
\left(1+\frac{3 T}{2 m_i }
                  + \frac{ 15  T^2}{8 m_i^2}+...
\right)
\\
p_i  & = & n_i T
\end{eqnarray}
We note that $\epsilon_i$ and $p_i$ are written in term of $n_i$ and T.
Therefore we can work
 with the variables $T, \mu_B,n_{\Lambda},n_{\bar{\Lambda}},etc$,
rather than $T,\mu_B,\mu_{\Lambda},\mu_{\bar{\Lambda}}, etc$.
The time dependence of $n_{\Lambda},n_{\bar{\Lambda}}, etc$ is known 
as discussed already.
So the modified equation of state can be computed from $t$, $T$ and 
$\mu_B$.

In figure~\ref{fig:sep},
we compare the behavior of $T$ and $\mu_B$ as function of $t$, 
obtained from the
hydrodynamical equations using the  modified equation of
state and the unmodified one. 
\begin{figure}[htbp]
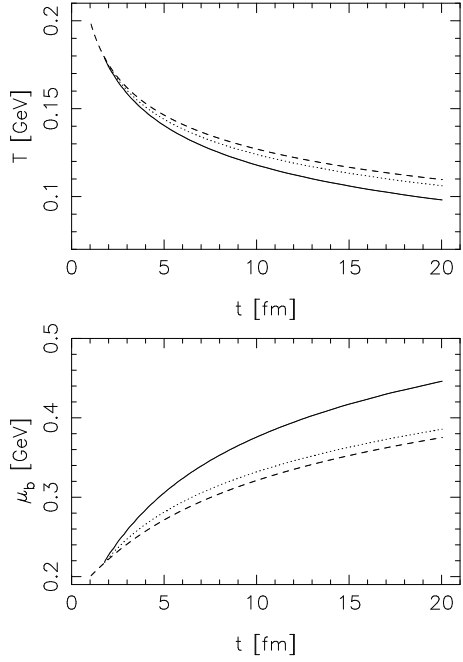

\begin{center}
\epsfig{file=T.ps,height=6.cm,angle=-90}\hfill
\epsfig{file=ub.ps,height=6.cm,angle=-90}
\caption{  $\mu_B$ and T as function of
  time in the case where all particles
have simultaneous freeze-outs (dashed line) and (I) all strange particles
in basic multiplets
 make an early chemical freeze-out (continuous line), (II)
 all strange particles except $K$ and $K^*$'s
 make an early chemical freeze-out (dotted line).}
\label{fig:sep}
\end{center}
\end{figure}
We see that
 if the chemical and thermal freeze-out
temperatures are very different
 or if many particle species make an early chemical freeze out,
 the thermal freeze out time is quite affected.
  Therefore
 it is important
to take into account the effect of the
early chemical freeze-out on the equation of state to make predictions
for observables which depend on thermal freeze-out volumes (which are related to 
the thermal freeze out time), 
for example
 particle abundances and  eventually particle correlations.

If the 
chemical and thermal freeze-out
temperatures are not very different  or if few
particle
species make the early freeze out,
one can proceed as follows.
One can use an unmodified equation of state in a
hydrodynamical code and to account for early chemical freeze-out of species
$i$, when the number of type $i$ particles was fixed,
use a modified Cooper-Frye formula
\begin{equation}
\frac{E d^3N_{i}}{dp^3}= 
\frac{N_{i}(T_{ch.f.})}{N_{i}(T_{th.f.})} \times
\int_{S_{th.f.}} d\sigma_{\mu} p^{\mu} f(x,p).
\end{equation}
The second factor on the right hand side is the usual one and it
gives the shape of the spectrum at thermal freeze-out, the first factor
is a normalizing term introduced such that upon
integration on momentum $p$, the number of particles of type $i$ is $N_i(T_{ch.f})$.
As an illustration, using
 HYLANDER-PLUS 
\cite{nel}, we show in figure~\ref{fig:hylander} 
 that both the shapes of $m_\perp$ spectra and
abundances can be reproduced for $T_{ch.f.}=176$ MeV and
 $T_{th.f.}=139$ MeV, while simultaneous freeze-outs at 
$T_{ch.f.}=T_{th.f.}=139$ MeV would yield
the correct shapes but too few particles.
\begin{figure}[htbp]
\begin{center}
\epsfig{file=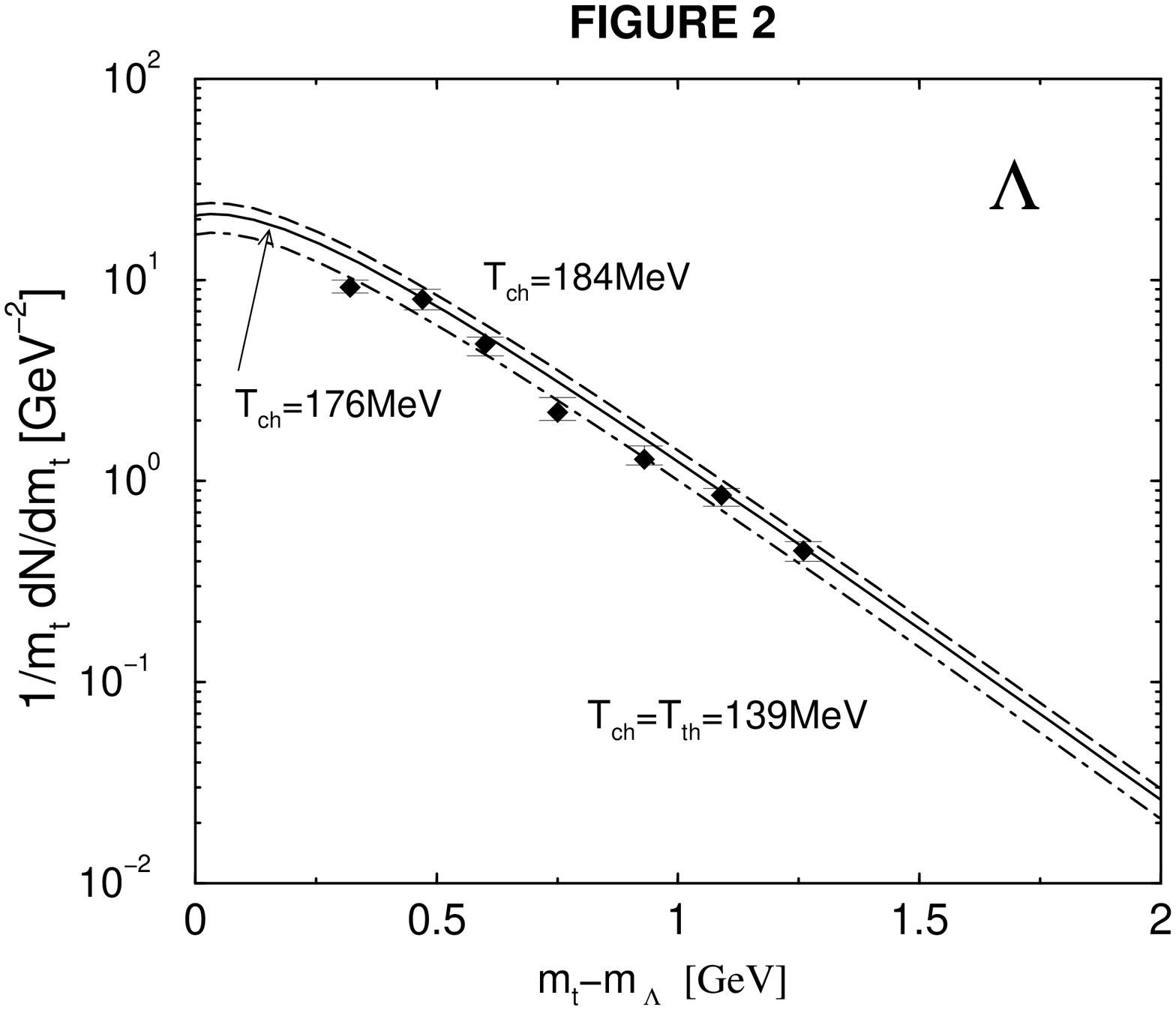,height=5.5cm,angle=-90}\hfill
\epsfig{file=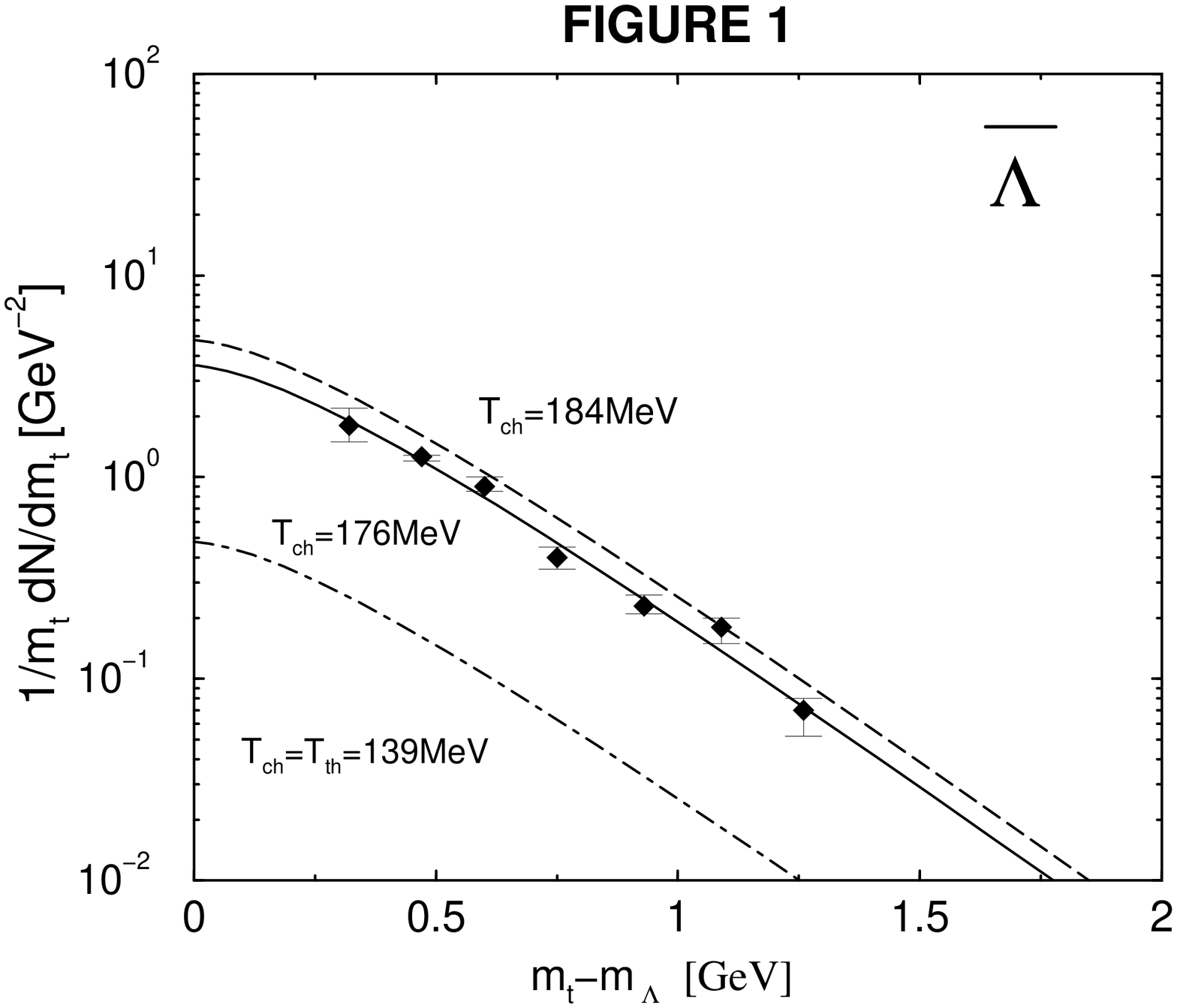,height=5.5cm,angle=-90}\\
\epsfig{file=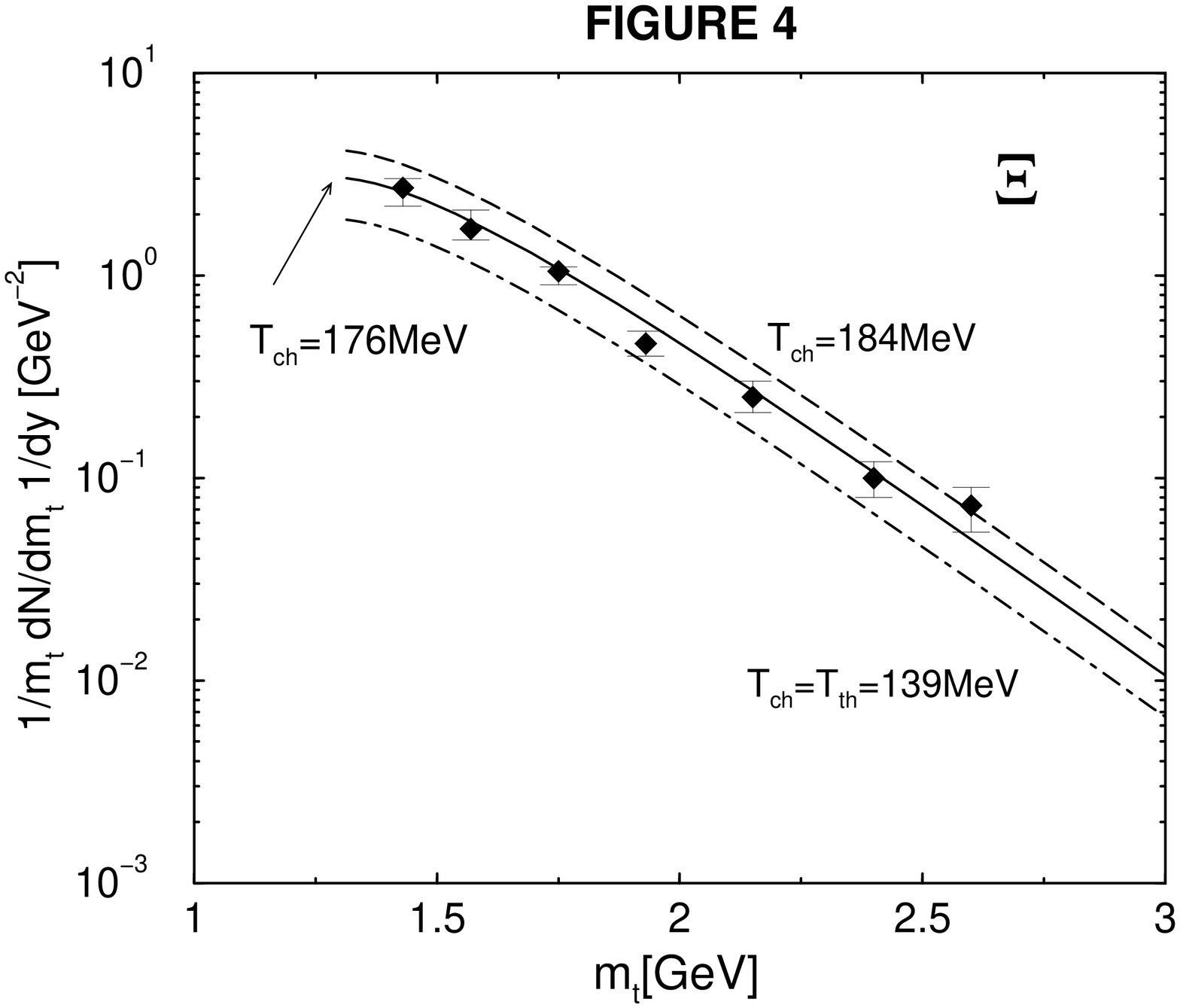,height=5.5cm,angle=-90}\hfill
\epsfig{file=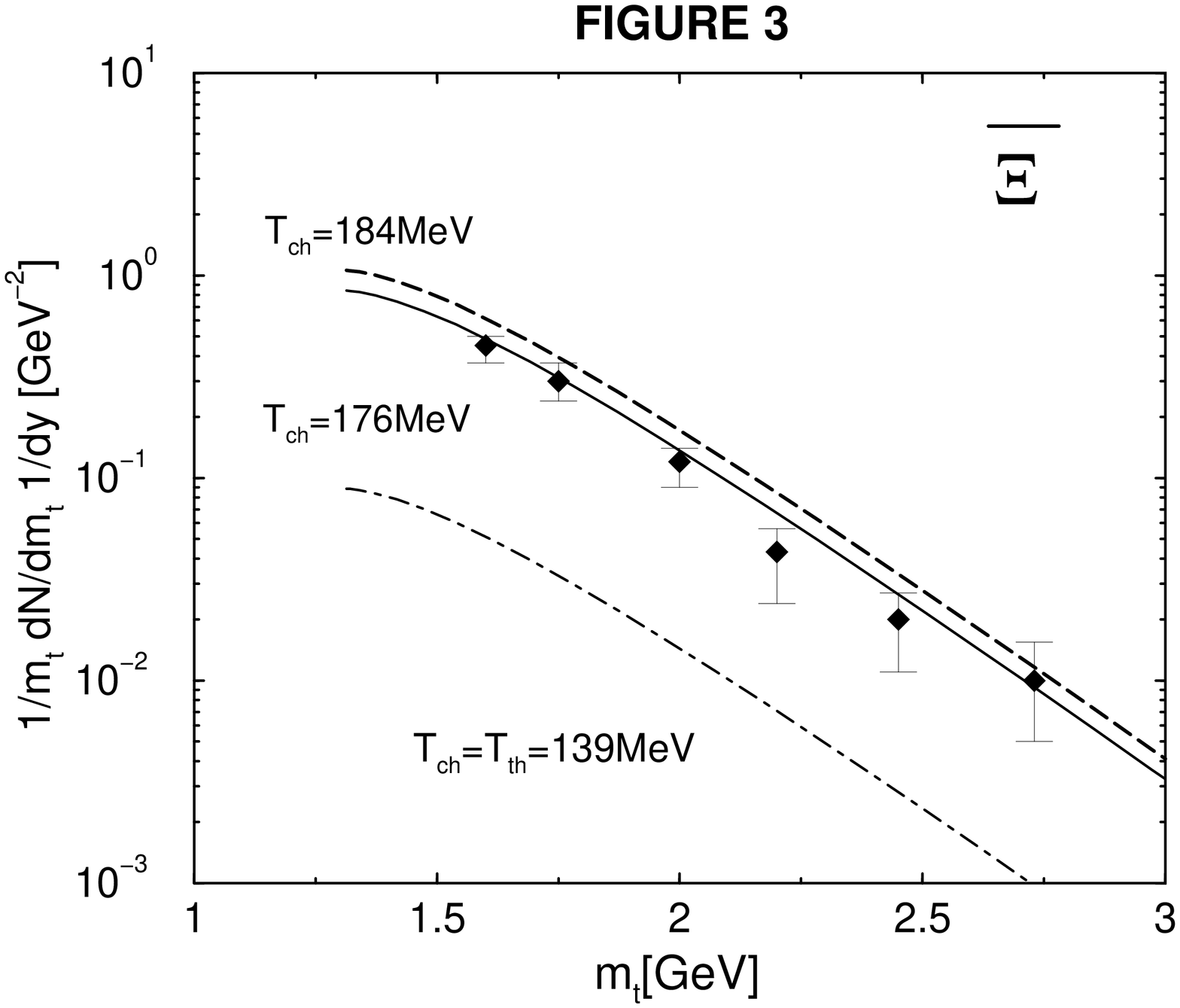,height=5.5cm,angle=-90}
\caption{  $m_{\perp}$ spectra:  NA49 data and
HYLANDER-PLUS results. Dashed-dotted curves:
single freeze out at $T_{f.out}=139 MeV$. Solid curves (resp. dashed): 
chemical freeze out at 
$T_{f.out.\,ch.}=$ 176 (resp. 184) MeV and thermal freeze out at
$T_{f.out.\,th.}=$ 139 MeV (see text). 
\label{fig:hylander}}
\end{center}
\end{figure}

Hirano and Tsuda \cite{hi} confirmed the importance of including
separate freeze outs in hydrodynamical codes, in particular they studied 
 elliptic flow and HBT radii at RHIC. This is also consistent with results
in \cite{teaney}.

\section{4. Continuous emission}

\subsection{4.1 Formalism and modified fluid evolution}

In this section, we present a possible way out of the second problem mentioned above.
In colaboration with
 Y.Hama and T.Kodama, I made a description of particle emission
 \cite{Grassi95a,Grassi96a} which incorporates the fact that
they, at each instant and each location, have a certain probability
to escape without collision from the dense medium
(said in the same terms as above: there exists a region in spacetime
for the last collisions of each particle).
So the distribution function of the system in expansion has two terms
$f_{free}$, representing particles that made their last collision already, and
 $f_{int}$, corresponding to the particles that are still interacting
\begin{equation}
f(x,p)=f_{free}(x,p)+f_{int}(x,p)
\end{equation}
The formula for the free particle spectra is given by
\begin{equation}
E d^3N/dp^3=\int d^4x\, D_{\mu} [p^{\mu} f_{free}(x,p)]
\label{eq:CEM}
\end{equation}
(neglecting particles that are initially free; note that if it were not the case, the use of hydrodynamics would not be possible).
$D_{\mu}$ indicates a four-divergence in general coordinates.
This integral must be evaluated for the whole spacetime occupied by the fluid.

This way, we see that the spectra contains information about the whole fluid history and not just the time when it is very diluted.
(This formula reduces to the Cooper-Frye formula
 (\ref{eq:CF}) in an adequate limit).

We can write
\begin{equation}
f_{free}={\cal P} f={\cal P}/(1 -{\cal P}) f_{int}.
\label{eq:free}
\end{equation}
${\cal P}(x,p)=f_{free}/f$,  the fraction of free  particles, 
can be identified with the probability
that a  particle of
 momentum $p$ escapes from $x$ without
 collisions, so to compute this quantity
we use the  Glauber formula
\begin{equation}
{\cal P}(x,p)=exp \left( - \int_t^{\infty} n(x') \sigma v_{rel} dt' \right).
\label{eq:glauber}
\end{equation}

We suppose also that all interacting particles are thermalized, so
\begin{eqnarray}
f_{int}(x,p) & = & f_{th}(x,p)= g/(2 \pi)^3  \nonumber \\
 & \times & 1/  \{\exp[(p.u(x)-\mu(x))/T(x)]\pm 1\}, 
\label{eq:termalizada}
\end{eqnarray}
where
$u^{\mu}$ is the fluid velocity,
 $\mu$ its chemical potential
($\mu=\mu_B B+\mu_S S$ with $B$ the baryonic number of the hadron,
$S$, its strangeness and $\mu_B$ and $\mu_S$ the associated chemical potentials)
and
$T$ its temperature at $x$. 

To compute
 $u^{\mu}$, $\mu$  and  $T$, 
we must solve the equations of conservation of energy-momentum and baryonic number
\begin{eqnarray}
D_{\mu}T^{\mu\,\nu}&=&0\\
D_{\mu}(n_b u^{\mu})&=&0.
\end{eqnarray}

In the
 figure~\ref{fig:evolution}, examples of solution are given.
We can see that the fluid evolution with continuous emission is
 different from the usual case  without continuous emission. 
For example, and as expected, the temperature decreases faster
since free  particles
carry with them part of the  energy-momentum.

In
 principle we have all the  ingredients to compute
  (\ref{eq:CEM}).
However there exist two problems:
 1) numerically, in the equation~\ref{eq:free} we can have divergencies if
 ${\cal P}$ goes faster to 1 than $f_{int}$ goes to zero 2) 
the hypothesis that
 $f_{int}$ is
termalized (cf. eq.(\ref{eq:termalizada})) must loose its validity when
${\cal P}$ goes to 1. 
To avoid this problem,
we divide  space-time in eq.
(\ref{eq:CEM}), in two  regions: the first with  ${\cal P} > {\cal
  P_F}$ and the  second with  ${\cal P} \leq {\cal P_F}$, for some
reasonable value of $ {\cal P_F}$. 
Using Gauss theorem, the  second part 
 reduces to an  integral over the  surface ${\cal P} = {\cal P_F}$ 
(which depends on the particle momentum)
\begin{eqnarray}
I_1 & = & \int_{{\cal P}={\cal P_F}} d\sigma_{\mu} p^{\mu} f_{free}\\
    & = & \frac{{\cal P_F}}{1-{\cal P_F}}
\int_{{\cal P}={\cal P_F}} d\sigma_{\mu} p^{\mu} f_{th}
\end{eqnarray}
There is still a certain fraction of interacting particles,
 $1-{\cal P_F}$ of the total,
on this  surface,
it is these  particles which in principle turn free
in the region
${\cal P} > {\cal P_F}$. To count them, we suppose that they are rather diluted
 (i.e.  ${\cal P_F}$ is rather large) and we can apply a  Cooper-Frye 
formula for them
\begin{eqnarray}
I_2 & \sim &  \int_{{\cal P}={\cal P_F}} d\sigma_{\mu} p^{\mu} f_{int}\\
    & = &  \int_{{\cal P}={\cal P_F}} d\sigma_{\mu} p^{\mu} f_{th}
\end{eqnarray}
So finally
\begin{equation}
E\frac{d^3N}{dp^3} =I_1+I_2 \sim \frac{1}{1-{\cal P_F}} 
\int_{{\cal P}={\cal P_F}} d\sigma_{\mu} p^{\mu} f_{th}.
\label{eq:CEMA}
\end{equation}
It is this formula which is used below,
with ${\cal P_F}=0.5$
(but we tested the effect of changing this value)
 and  coordinates adequate for the
geometry of the  problem.
It is similar to a  Cooper-Frye 
 formula (\ref{eq:CF}), however one must note that
the   condition ${\cal P}={\cal P_F}$ depends not only on the localization of 
a  particle but its  momentum, which as we will see has interesting  
consequences. 

\begin{figure}[!b]
\begin{center}
\epsfig{file=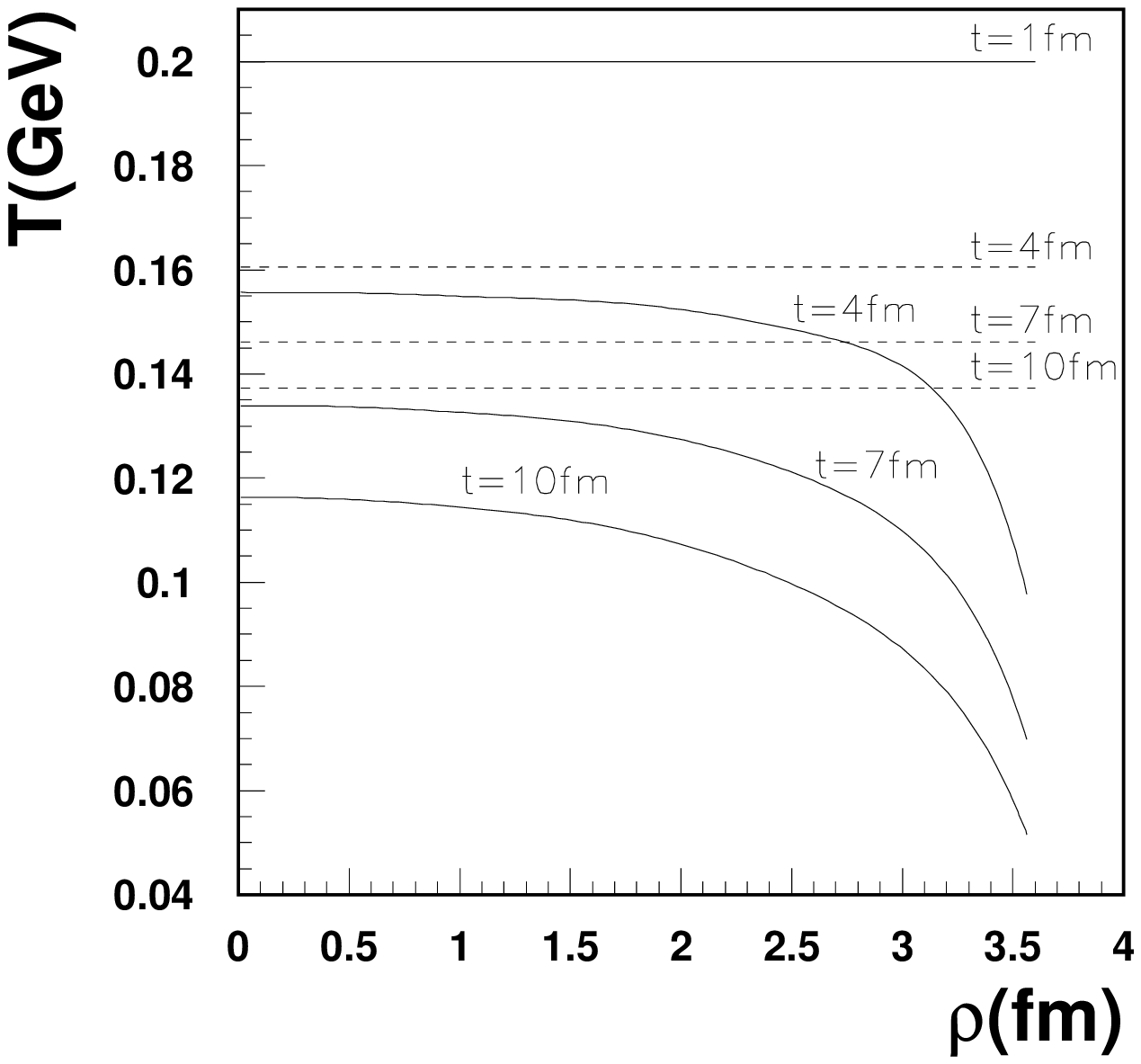,height=6.5cm}\\
\epsfig{file=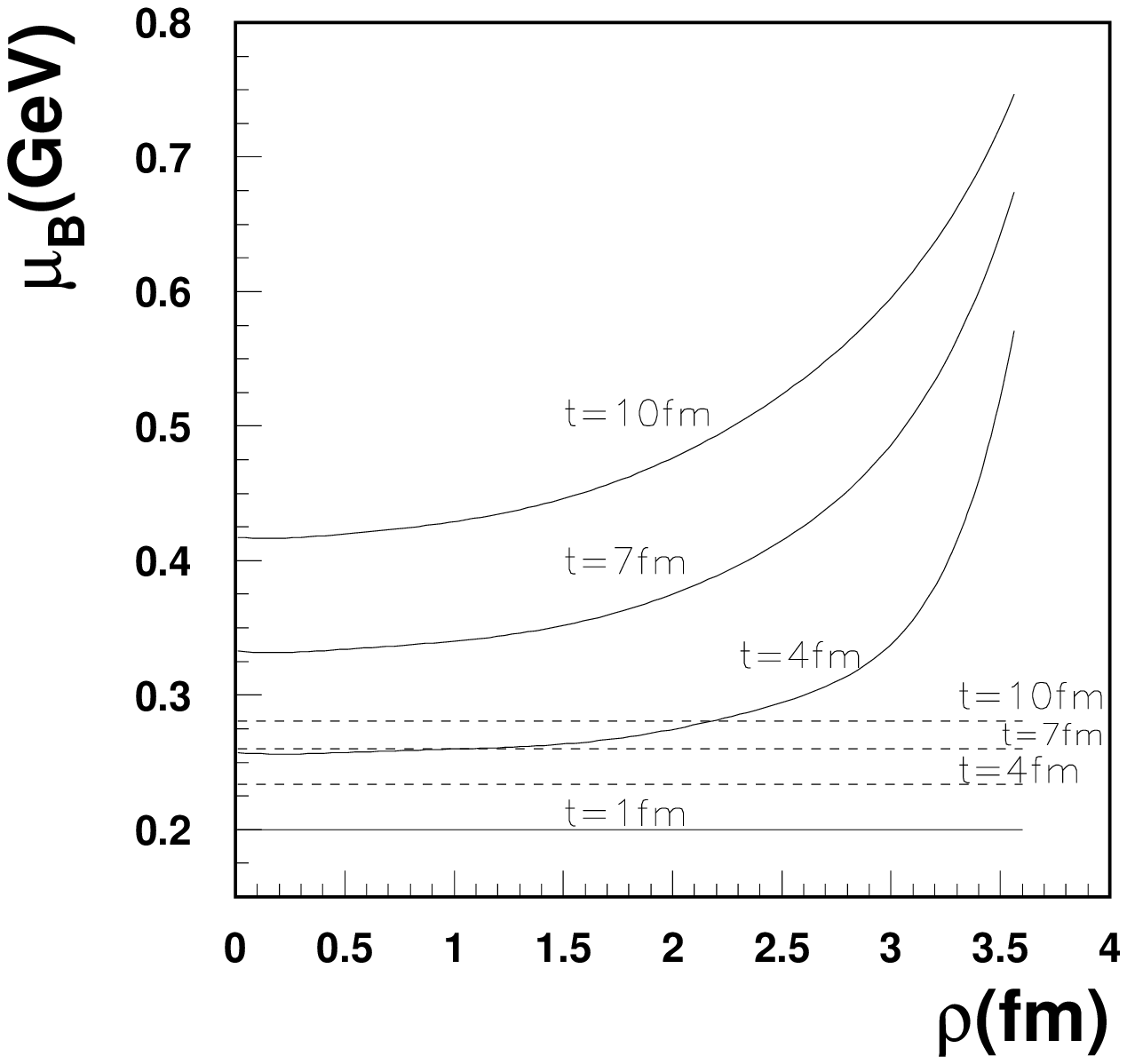,height=6.5cm}
\end{center}
\end{figure}
\begin{figure}[htbp]
\begin{center}
\epsfig{file=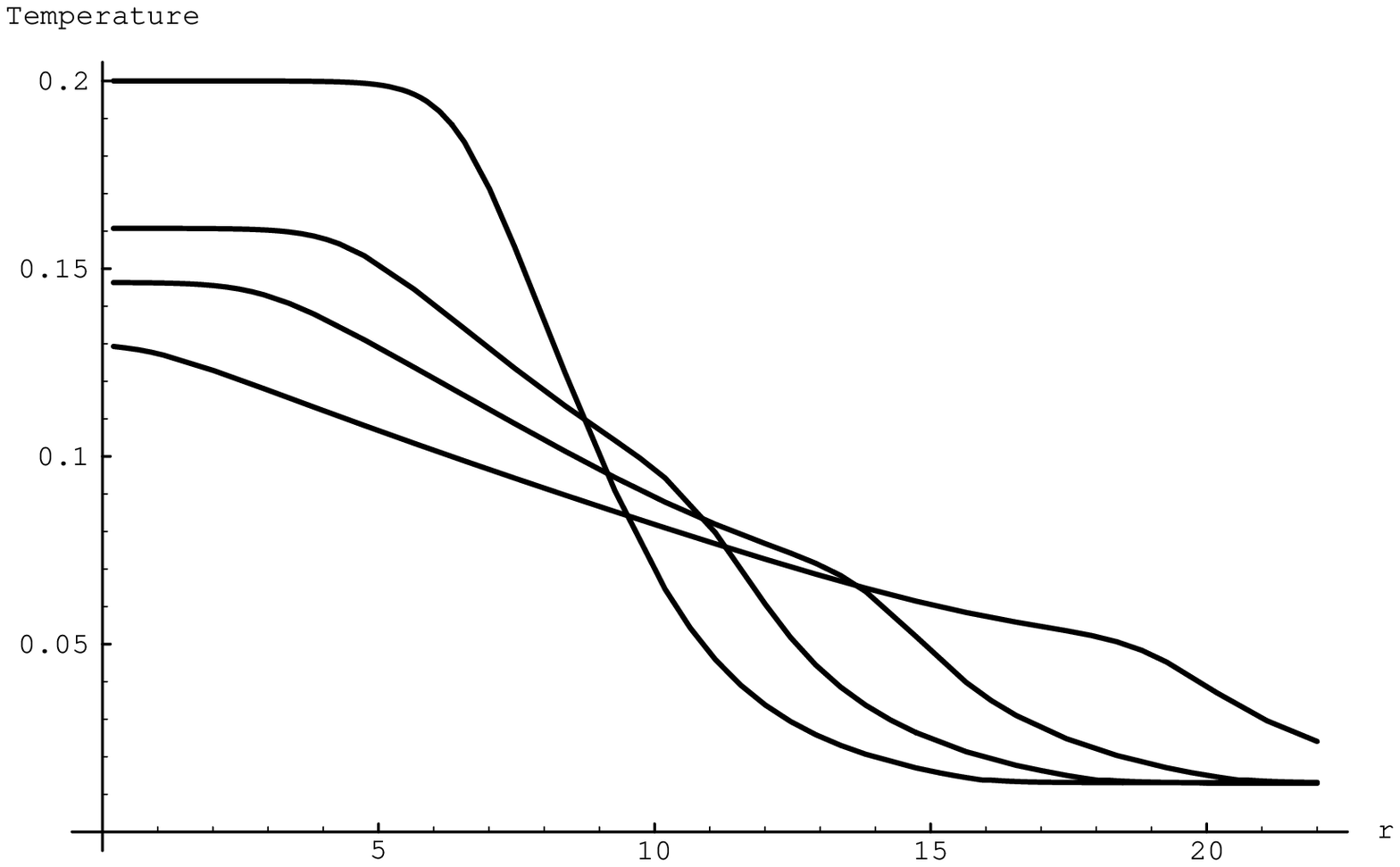,height=5.cm}\\
\epsfig{file=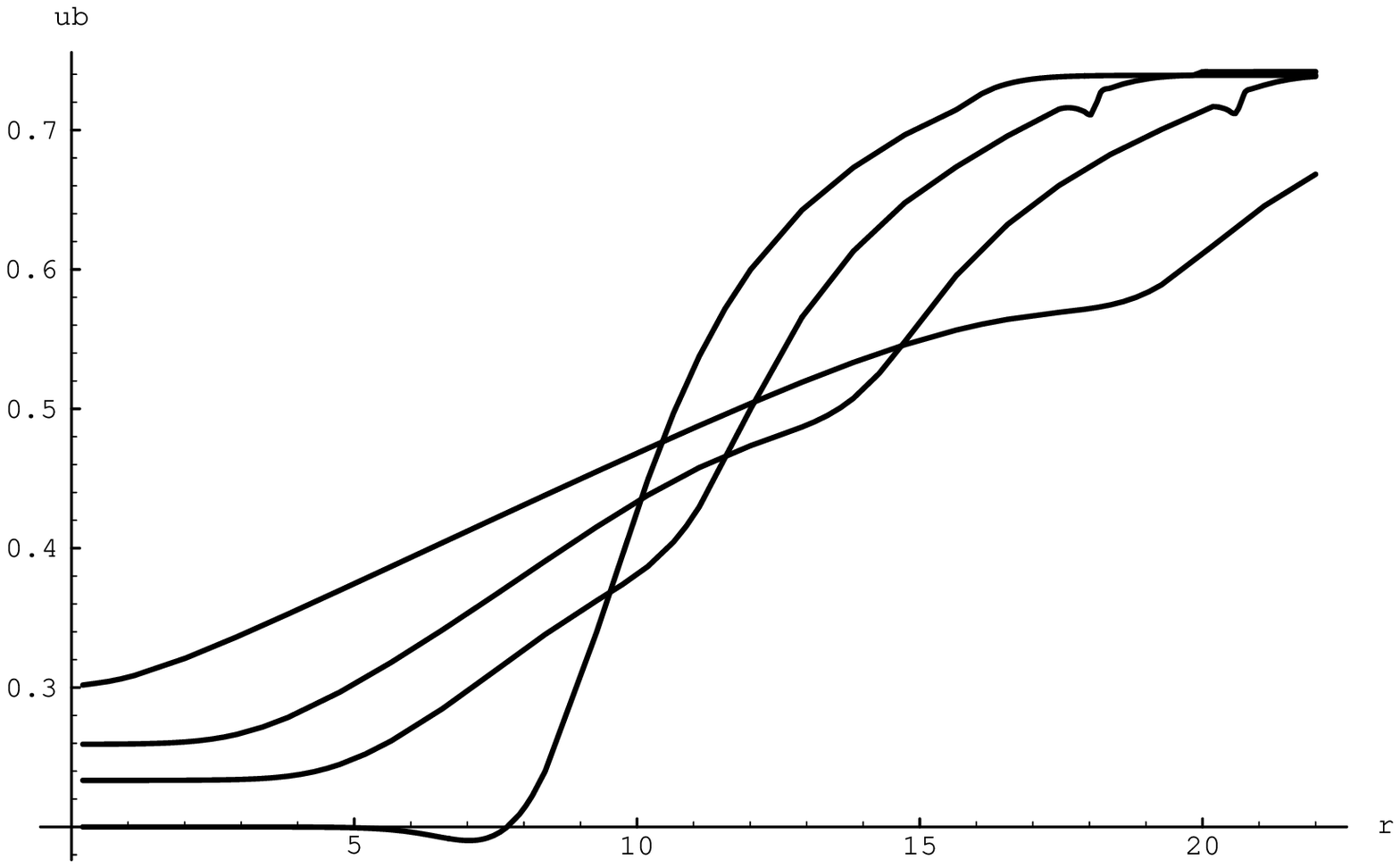,height=5.cm}\\
\epsfig{file=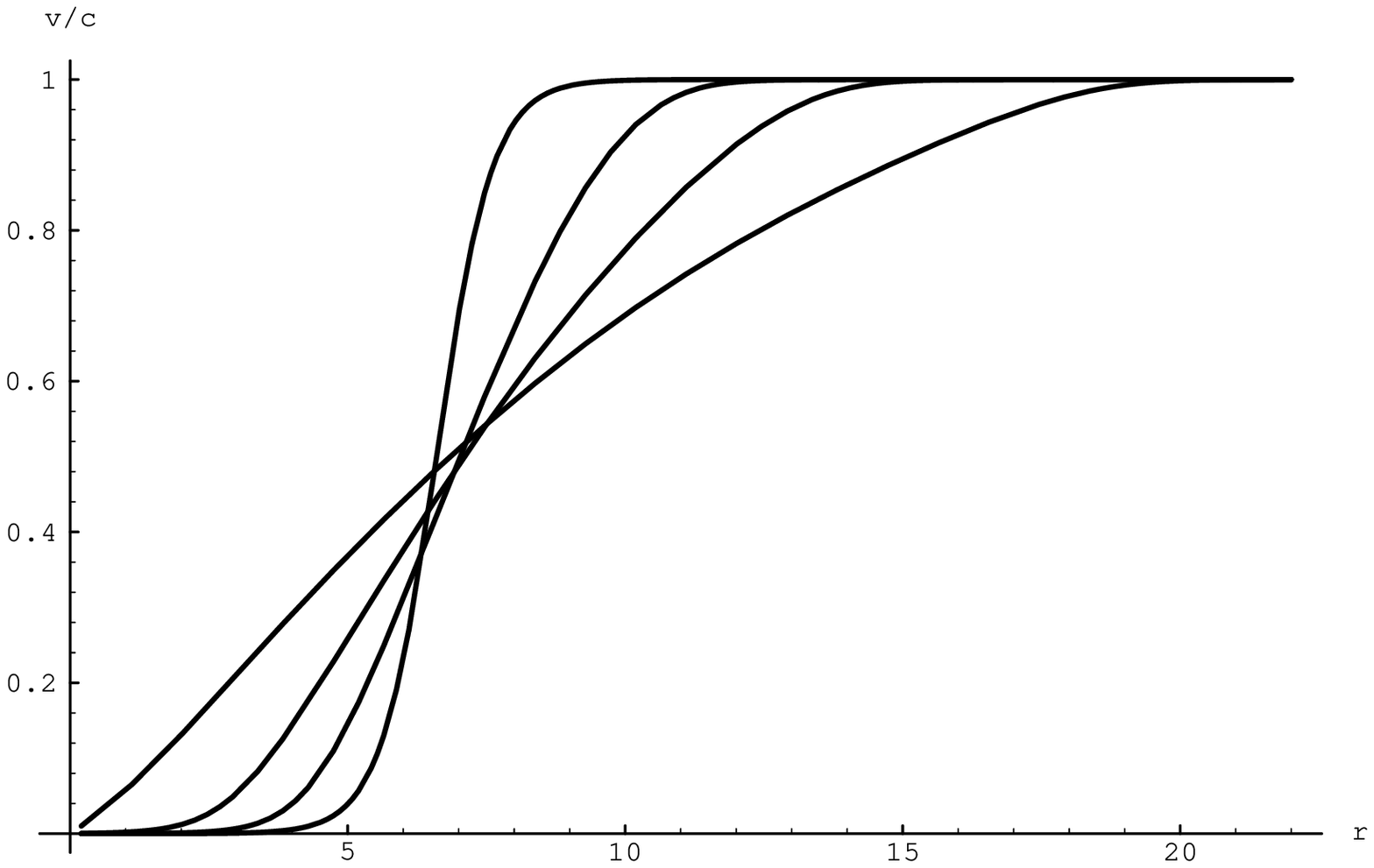,height=5.cm}
\caption[Fluid evolution]{  Fluid evolution
 supposing
longitudinal boost invarience\cite{bj}. 
Top two: for a light projectile such as S, $T$ and $\mu_B$ as function of radius; 
as a first
  approximation transverse expansion is neglected.
( Solid lines correspond to a fluid with
  continuous emission,  dashed line lines
 to a fluid without continuous emission, i.e. the usual case.)
Bottom three: for a heavy projectile such as  Pb, $T$, $\mu_B$ and fluid velocity as function of radius;
  transverse expansion is included (times: 1,4,7,10 fm).
}
\label{fig:evolution}
\end{center}
\end{figure}

\subsection{4.2 Comparison of the continuous emission and freeze out scenarios}

In this section, I compare the interpretation of experimental data
in both   models.\\

{\bf a. Strange particle ratios}\\
We saw above that for the freeze out mechanism, strange particle ratios give information about chemical freeze out.
Now we see how to interpret these data within the continuous emission sceanrio \cite{Grassi96c,Grassi97,Grassi98,Grassi99a,Grassi99b}.
In this case, the only parameters are the initial conditions
$T_0$ e $\mu_{b\,0}$ and a value that we suppose average for
 $\gamma$.
We therefore fix a set of them, solve the
 equations of 
hydrodinamics with continuous emission, compute and integrate in  $p_{\perp}$   the spectra given by (\ref{eq:CEMA}) 
for each type of particles (we include also the decays of the various types of particles in one another).
In a way similar to freeze out
but now with   initial values instead of  freeze
out values, we get in
 figure~\ref{fig:janelaWA85},  
a window of  initial   conditions which permits to  reproduce the various 
 experimental ratios. (We also look at other ratios than those shown
 in the  figure, tested the effect
of changes in the equation of state, cross section,
  initial time, type of experimental cutoff.) 

\begin{figure}[htbp]
\begin{center}
\epsfig{file=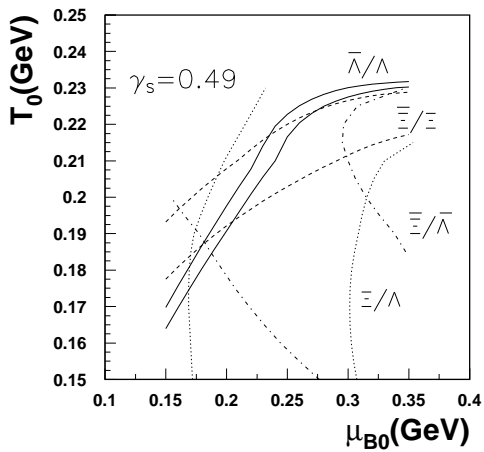,height=6.5cm}
\end{center}
\caption{ Window in  initial conditions
 allowing to reproduce WA85 data, for an equation of state with excluded volume
corrections~\cite{Grassi98}.}
\label{fig:janelaWA85}
\end{figure}
We see therefore that the 
initial
 conditions  necessary to  reproduce the WA85 data are
\begin{equation}
T_0\sim\mu_{b\,0}\sim 200\,MeV
\end{equation}
These values may seem high for the existence of a hadronic phase,
lattice gauge QCD simulations 
seem to 
indicate values smaller for the quark-hadron
transition.
Here we can note that 1) values of  QCD on the lattice are still evolving
 (problems exist to incorporate quarks with intermediate mass,
 include $\mu_b \neq 0$, etc. 2) Our own model is still being improved
and we know for example that the equation of state affects the  localization
of the window in initial conditions compatible with data.

In figure~\ref{fig:janelaWA97}, the same kind of analysis was done for the data  WA97.
\begin{figure}[htbp]
\begin{center}
\epsfig{file=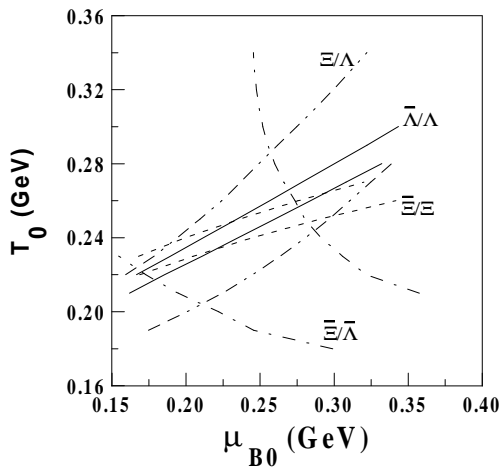,height=6.5cm}
\end{center}
\caption{ Window in initial conditions  reproducing
WA97 data in the
 case of an equation of state without
 excluded volume corrections. With volume
corrections, the window is lower \cite{Grassi99b}.}
\label{fig:janelaWA97}
\end{figure}
With the reservation
in the caption of the figure, we see that the initial
conditions are not very different from the one above.

We can therefore conclude that the interpretation
of data on particle ratios lead to totally different  information
according to the emission model used. For the freeze out model, we get information about chemical freeze out while for continuous emission,
we learn about the initial conditions. \\

{\bf b. Transverse mass spectra}\\
For
freeze out, we saw that these spectra tell us about thermal freeze out (temperature and fluid velocity).
 Now for continuous emission let us see how to interpret these same spectra
 \cite{Grassi96c,Grassi97,Grassi98, Grassi99a,Grassi99b}.
In this case, the initial conditions were already determined from
strange particle ratios, they cannot be changed and must be used to compute the spectra as well.
For example, in figure~\ref{fig:espectros}, various spectra are shown, 
assuming $T_0=\mu_{b\,0}=200$ MeV, and compared with experimental data.

\begin{figure}[htbp]
\epsfig{file=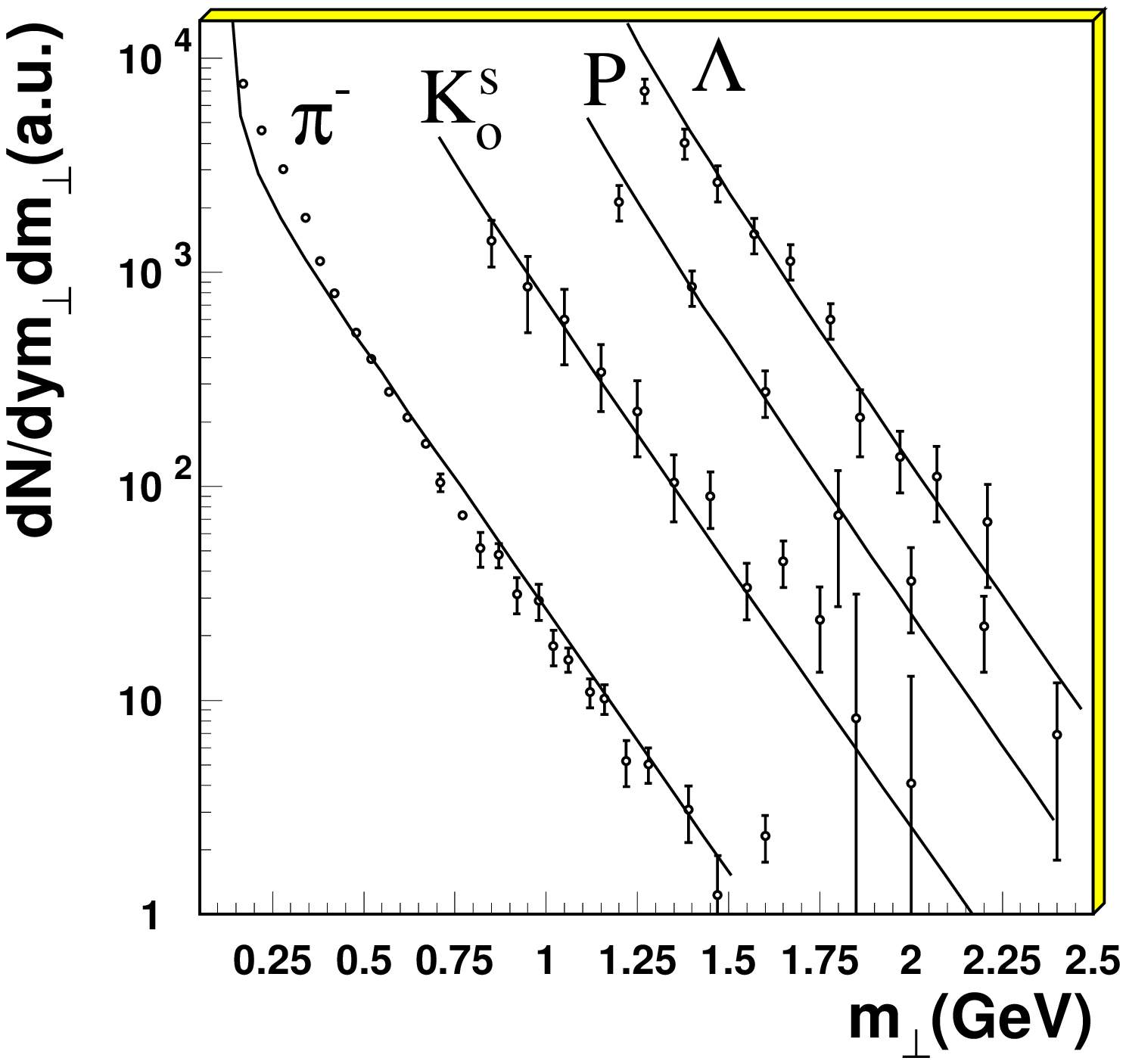,height=7.cm} 
\epsfig{file=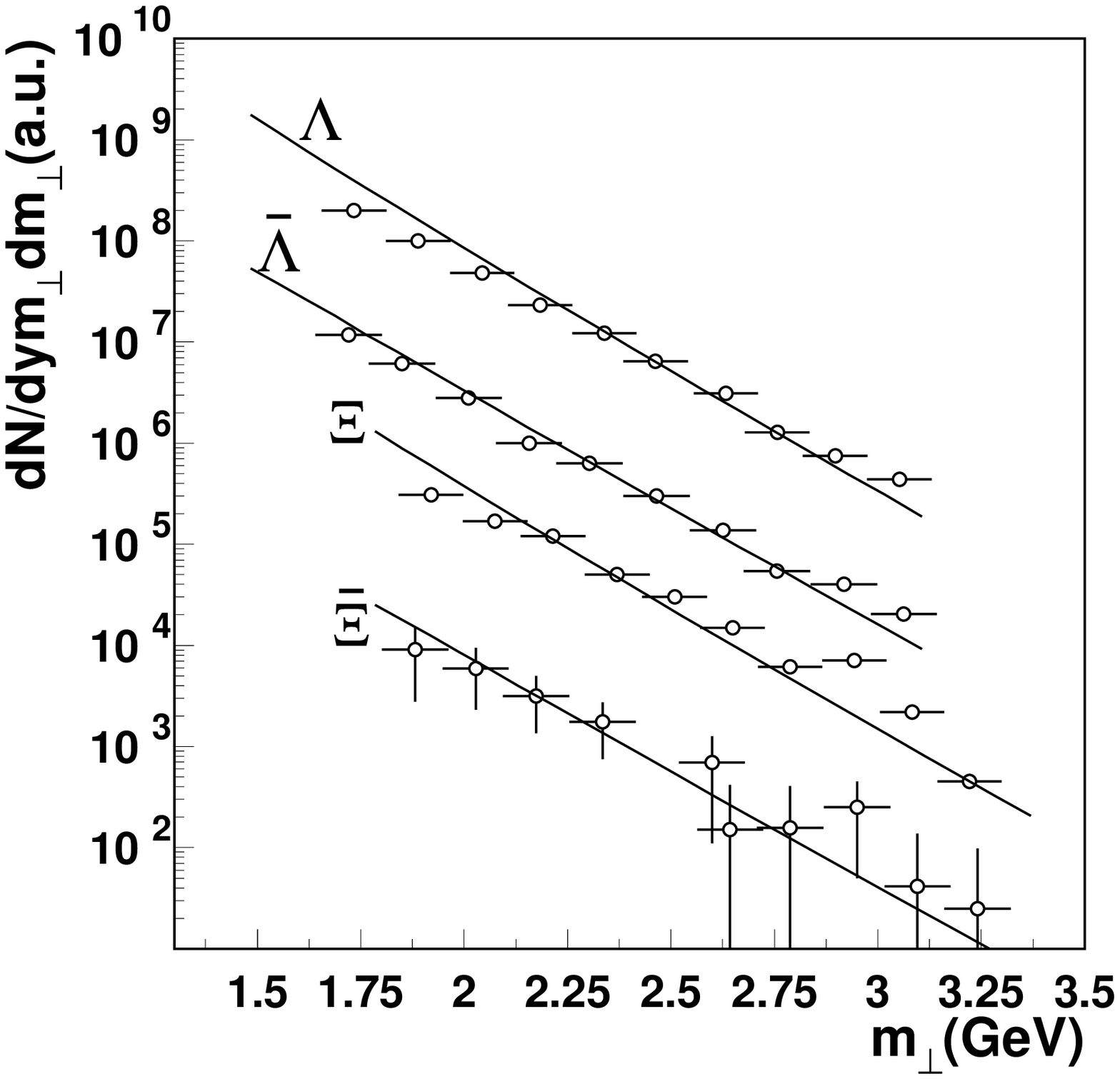,height=6.5cm}
\epsfig{file=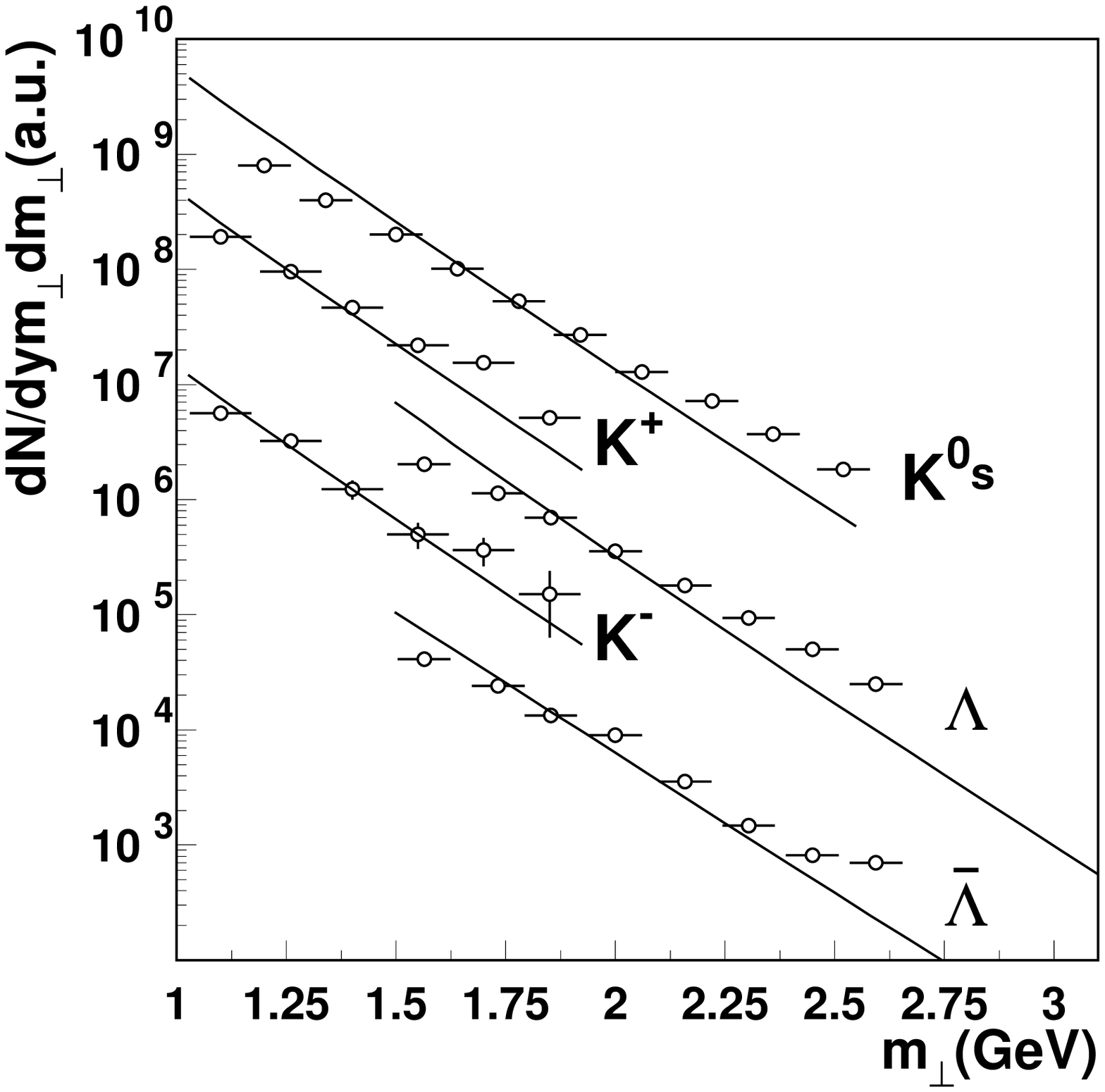,height=6.5cm}
\caption{ Using longitudinal boost invariance and no transverse expansion, we compare 
our predictions, top to bottom, with NA35 data   (S+S, all rapidity),
with  WA94 data (S+S, midrapidity, large transverse momenta)
and with WA85 data (S+W, midrapidity, large transverse momenta)
\cite{Grassi99a}.}
\label{fig:espectros}
\end{figure}
This comparison should not be considered as a fit but as a test of the possibility of
 interpreting various types of data in a self-consistent way with continuous emission 
(in particular note that our calculations were done assuming longitudinal
boost invariance). 

We learn various information from this comparison.
In these figures,
we do not take into account transverse expansion. 
With
the S+S NA35 data, we note that the heavy 
 particles and high transverse momentum  pions 
have similar inverse inclinations  $T_0 \sim 200$ MeV. 
The particles heavier than pions, due to 
termal suppression, are mainly emitted
early when the temperature is $\sim
T_0$. For lower temperatures, there are still emitted (and more easily due to matter dilution)
but their  densities are quite smaller
 (this is what is called thermal suppression) and their contributions as well.
The high transverse momentum  pions 
have large velocity and
 (if not too far away from the outer fluid surface) escape without collision
earlier than pions at the same place but with smaller velocity,
so these high transverse momentum  pions
also escape at $\sim T_0$.
Pions are small mass particles and  are little affected by thermal suppression.
This way, they can escape in significative number at various temperatures.
This is reflected by their spectra, precisely its curvature.
(In our calculation, decays into pions are not included, this would
fill the small transverse momentum region and improve the agreement with experimental data.)

The S+S WA94 data
also indicate that continuous emission is compatible with data.
Finally, the S+W WA85 data  seem to indicate
that perhaps somewhat different initial conditions  or a little  of transverse expansion might be necessary to reproduce data.

Our   calculations including transverse expansion
indicate that little transverse expansion is compatible with data for light 
projectile.
This is understandable noting that the effective temperature of spectra are 
already of order the initial temperature $\sim$ 200 MeV.
In the case of heavy projectile, the situation is different.
The various types of particles have different temperatures, all well above 200 MeV. In this case, we must include transverse expansion to get consistency with data.
An example is shown in
 figure~\ref{fig:ptwa97} with the same values of
 $T_0$ and $\mu_{b,0}$ than previously, 
200 MeV.
\begin{figure}[htbp]
\begin{center}
\epsfig{file=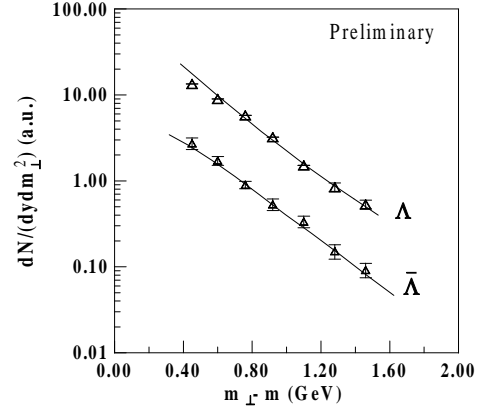,height=6.5cm}
\end{center}
\caption{ WA97 data and  example of comparison with
continuous  emission with transverse expansion
   (not a least square fit) for $T_0=\mu_{b.0}=200$ MeV \cite{Grassi99b}.} 
\label{fig:ptwa97}
\end{figure}
\\

{\bf c. The case of the $\Omega$}\\
The effective temperatures in the case of heavy projectiles, particularly for the $\Omega$, have attracted a lot of attention for the following reason.
In the usual hydrodynamics with freeze out,
as we saw, it is expected that effective temperatures increase with mass.
In this context it was difficult to understand why the
 $\Omega$ had an effective temperature much lower than other particles in
figure~\ref{fig:temp}.
\begin{figure}[htbp]
\begin{center}
\epsfig{file=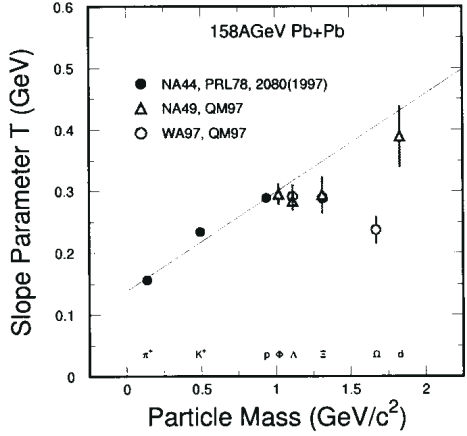,height=6.5cm}
\end{center}
\caption{ Compilation of experimental data
 on effective temperatures
 in the case of heavy projectile and  predictions for usual
  hydrodynamics \cite{va98}. (The various experiments have different
 rapidity  and transverse momentum cutoffs).
In  general, deuteron is left outside the hydrodynamical analysis:
 it is little bounded and should form later in the fluid evolution from
   coalescence of a neutron and 
  a proton with similar moment.}.
\label{fig:temp}
\end{figure}
A possible explaination within hydrodynamics with
separate freeze outs, is that the
 $\Omega$ made its chemical and thermal freeze out together, early.
Van Hecke et al.~\cite{va98} argued that this
 is reasonable since it is expected that  $\Omega$  has a small cross section
(because there is no channel for  $\Omega \pi 
\rightarrow resonance \rightarrow \Omega \pi$)
and
showed that the microscopical model RQMD
can reproduce the data.
\begin{figure}[htbp]
\begin{center}
\epsfig{file=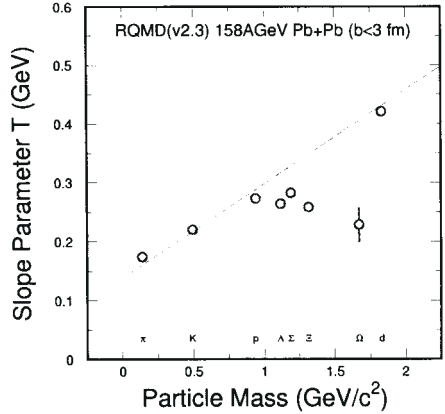,height=6.cm}
\end{center}
\caption{Predictions from  RQMD \cite{va98} in the same experimental situation as the previous figure.}
\label{fig:tempr}
\end{figure}

In our model originally we had used the same value of the cross sections
to compute the escape probability for the various types of particles. 
This is not expected and indeed in this case, continuous 
emission,  
does not lead to good predictions as shown in figure~\ref{fig:tempec}a.
\begin{figure}[htbp]
\begin{center}
\epsfig{file=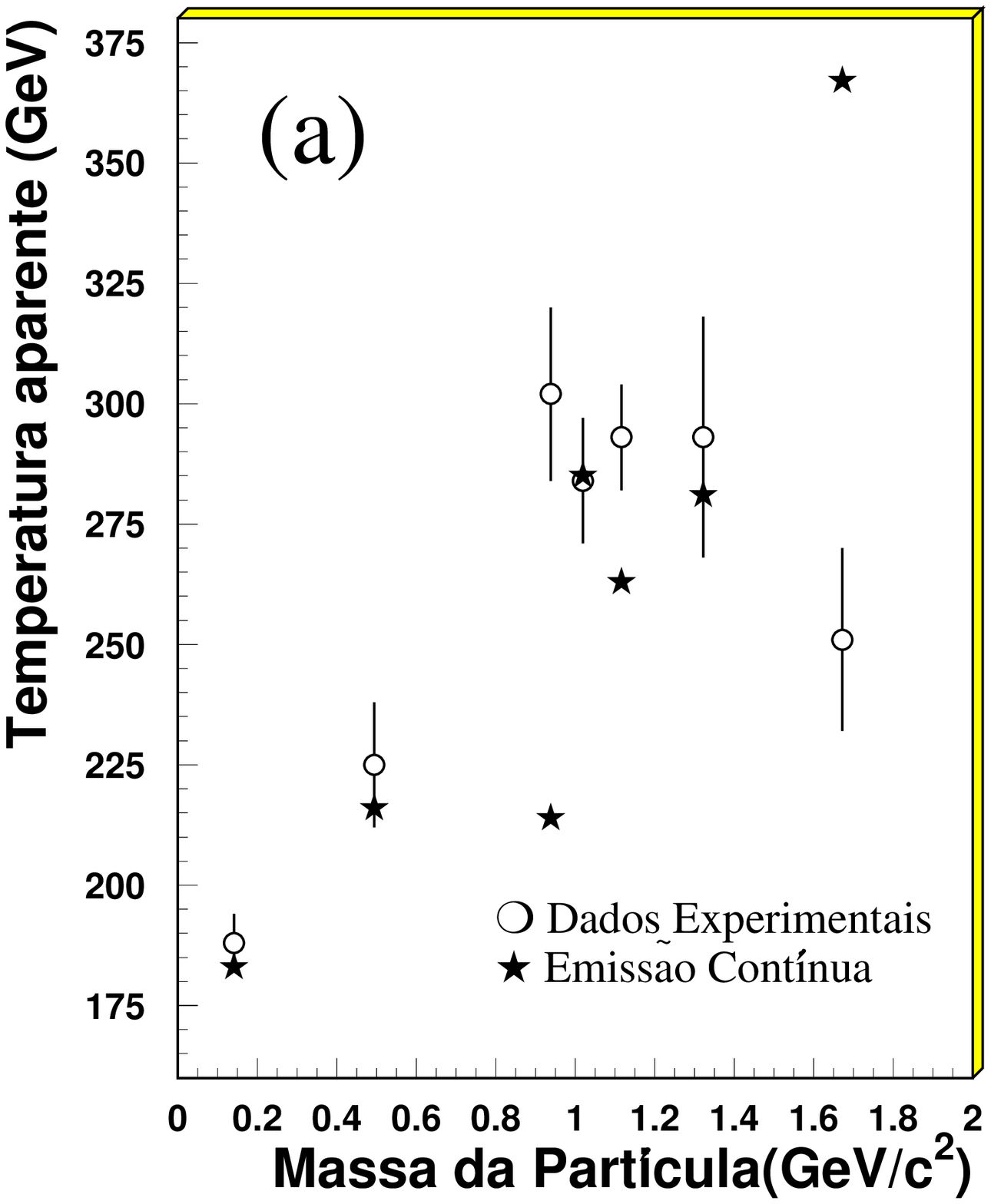,height=7.5cm}\hfill\epsfig{file=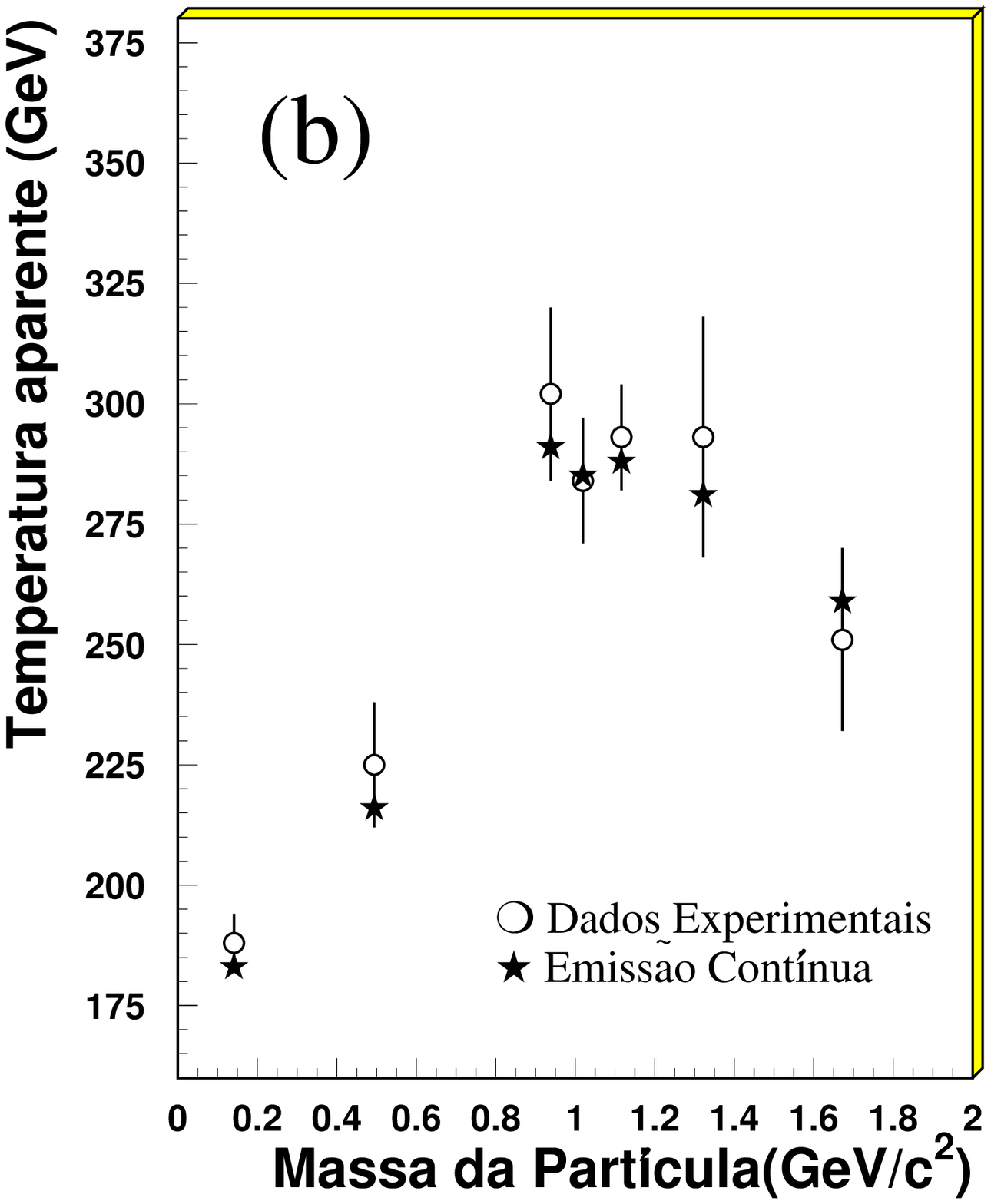,height=7.cm} 
\end{center}
\caption{ Comparison  of experimental data on effective  
temperatures
 in the case of a heavy  projectile  with  predictions from continuous emission. Top: all cross section supposed equal. Bottom: more realistic cross section
values
 (cf. text) 
\cite{os99,Grassi99b}.}
\label{fig:tempec}
\end{figure}
Therefore in the spirit of microscopic models, we also show our predictions 
in  figure~\ref{fig:tempec}b, for continuous emission and the following
cross sections:
$\sigma_{\pi \pi}\sim <\sigma v_{rel}>_{\pi \pi}\sim 1 fm^2$,
$\sigma_{N \pi} \sim
<\sigma v_{rel}>_{N \pi} \sim 3/2 <\sigma v_{rel}>_{\pi \pi}$
(using additive  quark model estimate),
$\sigma_{\Lambda \pi} \sim
<\sigma v_{rel}>_{\Lambda \pi} \sim 1,2 <\sigma v_{rel}>_{\pi \pi}$
(using additive  quark model estimate \cite{ba98})
and
$\sigma_{\Omega \pi} \sim
<\sigma v_{rel}>_{\Omega \pi} \sim 1/2 <\sigma v_{rel}>_{N \pi}$
(using estimate in \cite{bravina98b}).
The predictions now are in agreement with data. However, the cross sections are poorly known and our results
 are  sensitive to their values.

Recently, with new data by NA49, WA97~\cite{WA97art} and NA49~\cite{NA49art}
came to the conclusion that in fact there is no need for early joint chemical 
and thermal freeze outs for the 
$\Omega$: all their spectra can be fitted with a simple hydro inspired model
as seen in figure~\ref{fig:omega}.
The previous difficulty for WA97 came from the fact~\cite{heinzart}
 that $\Omega$ was observed at high $p_{\perp}$. 
However now, STAR has problems \cite{starart}
to fit with a simple hydro inspired model
the $\Xi$ together with 
$\pi, p, K, \Lambda$ and would need to assume early joint chemical and thermal freeze outs for the $\Xi$, as shown in figure~\ref{fig:cascade}.
More recently, it has been noted \cite{na49omega}
by NA49,
that their conclusion depends on the parametrization used and
in \cite{na57omega}, 
NA57 argues that due to low statistics, it is not clear what conclusion can be drawn for 
the $\Omega$.
In \cite{fgsqm}, an attempt was made  to reproduce with a single  thermal 
freeze out temperature in a hydrodynamical code,
all transverse mass spectra 
at a given energy.
\begin{figure}[htbp]
\begin{center}
\epsfig{file=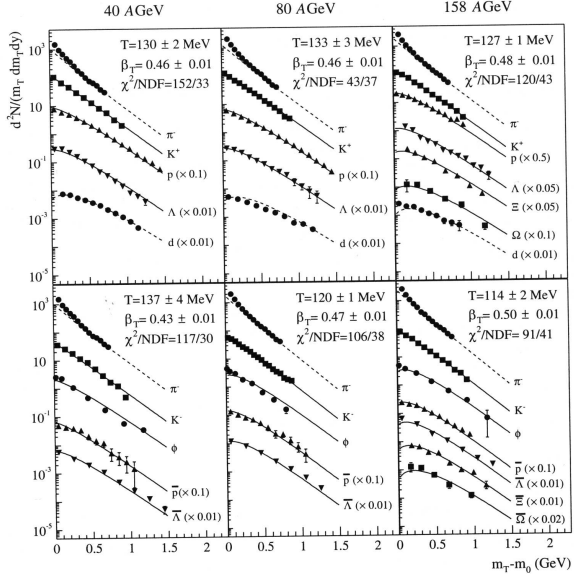,height=8.cm,angle=180}
\epsfig{file=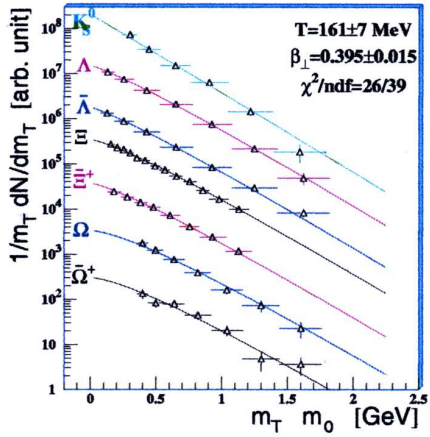,height=6.cm,angle=0}
\end{center}
\caption{Data and hydro inspired fit for all particles including $\Omega$, 
top, NA49 and bottom, NA57 \cite{WA97art,NA49art}.}
\label{fig:omega}
\end{figure}
\begin{figure}[htbp]
\begin{center}
\epsfig{file=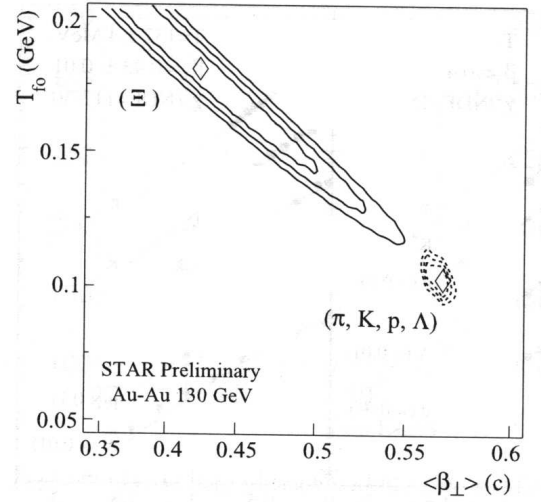,height=7.cm,angle=0}
\end{center}
\caption{A hydro fit to transverse mass spectra leads to high $T_{f.out}$
for $\Xi$ and lower for $\pi, p, K, \Lambda$ \cite{starart}.}
\label{fig:cascade}
\end{figure}
\\

{\bf d. Pion abundances}\\
We showed that
strange particle ratios can be reproduced by a model with chemical freeze out around 180 MeV. The abundances too can be reproduced. The problem is that the pion number is too low. This was noticed by
  Davison et al. \cite{Davidson92a}, as shown in their table~\ref{tab:davidson}
reproduced below

\begin{table}[htbp]
\caption{Comparison between NA35 data and previsions
  from a thermal model
\cite{Davidson92a}.}
\label{tab:davidson}
\begin{center}
\begin{tabular}{|l|lllllll|}
\hline
 & $K_S^0$ & $K^+$ & $K^-$ &  $\Lambda$ & $\overline{\Lambda}$ & ``p'´
 & $\pi^-$\\
\hline
Th.          &  10.7          &  14.2         & 7.15        & 8.2
               &  1.5           & 23.2          & 56.9               \\
Exp.     &  10.7  &  12.5 & 6.9 
               & 8.2    &  1.5   & 22.0  & 92.7
 \\
               &  $\pm$2.0  &  $\pm$0.4 & $\pm$0.4 
               &  $\pm$0.9  &  $\pm$0.4   & $\pm$2.5  & $\pm$4.5       \\
\hline
\end{tabular}
\end{center}
\end{table}

There are various ways to try to solve this problem.
\begin{enumerate}
\item It can be argued  (in a spirit similar to  Cleymans et al. \cite{cl93})
that strange particles do their chemical freeze our early 
around
 180 MeV and their thermal freeze out around
 140 MeV but the
 pions do their chemical and thermal freeze outs together around
 140 MeV. In fact
Ster et al. \cite{st99}
manage to reproduce data by  NA49, NA44 and WA98 on spectra 
   ({\em including normalization i.e. abundances}) for
 negatives, pions, kaons and protons  and
HBT radii (see next section) with temperature around
 140
MeV and null chemical potential for the pions.
On the other side,  Tom\'asik et al. \cite{to99}
say that for the same kind of objective, they need
temperatures of order  100 MeV and non zero chemical potential for the pions.
So it is not clear if to reproduce the pion abundance, 
it is necessary to modify the hydrodynamics with separate freeze out, 
including 
pions out of chemical equilibrium or not.
\item Gorenstein and colaborators \cite{ri97,ye97,ye99}
studied modifications of the  equation of  state, precisely they included
a smaller radius for the volume corrections of the pion. 
\item Letessier et al. made a  s\'erie of papers \cite{le93,letessier95}
arguing that the large pion abundance is indicative of the formation of a quark gluon plasma
hadronizing suddenly, with both strange and non-strange quarks out of chemical equilibrium~\cite{le98}.
\end{enumerate} 

Given the difficulty that freeze out models have with pions, it is interesting to compute abundances with continuous emission models.
In the table, results and  NA35 data from S+S are shown (data are selected at midrapidity).
\begin{table}[htbp]
\caption{Comparison between  experimental data, results for
continuous  emission at $T_0=\mu_{b,0}=200$ MeV and freeze out at
   $T_0=\mu_{b,0}=T_{f.out}=\mu_{b\,f.out}=200$ MeV.}
\begin{center}
\begin{tabular}{|c|l|l|l|} \hline
  & experimental value  &  continuous emission  & freeze out\\  \hline
$\Lambda$ & 1.26$\pm$0.22 & 0.96 & 0.92 \\
$\bar{\Lambda}$ & 0.44$\pm$0.16 & 0.29 & 0.46 \\
$p-\bar{p}$  & 3.2$\pm$1.0 & 3.12 & 1.32 \\
$h^-$     & 26$\pm$1 & 27 & 15.7 \\
$K^0_S$   & 1.3$\pm$0.22  & 1.23 & 1.06\\ \hline
\end{tabular}
\end{center}
\end{table}

It can be seen that in the continuous emission model, a larger number of pions is created.
In \cite{Grassi00a}, we related this increase to
the fact that entropy increases during the fluid expansion. 
This is due to the continuous process of separation into free and interacting components  
as well as continuous
 re-termalization of the fluid. 
In contrast, in the usual hydrodynamic model,
 entropy is  conserved and  related to the  
   number of  pions; so in this usual model,  observation of a large
 number of  pions
implies a large initial   entropy and is indicative of a 
   quark gluon plasma  \cite{le93,letessier95,le98}.
In our case, a large number of pions does not imply a large initial entropy 
and the existence of a plasma. 
It can be noted that the interpretation of data is quite
influenced by the choice of the particle emission model.\\

{\bf e. HBT}\\
Interferometry is a tool which permits  extracting information on the spacetime structure of the   particle emission source and is sensitive to the underlying dynamics.
Since pion emission is different in freeze out and continuous emission models,
it is interesting to compare their interferometry predictions (for a review see \cite{sandra}). 
This was done in  ref. \cite{Grassi00b}.

In this work, the formalism of continuous emission
\cite{Grassi95a,Grassi96a}
was extended to the computation of  correlation functions. 
Precisely, we computed
\begin{equation}
C(k_1,k_2)=C(q,K)=
1+\frac{|G(q,K)|^2}{G(k_1,k_1)G(k_2,k_2)}\;\;,  \label{cce}
\end{equation}
where $q^\mu =k_1^\mu -k_2^\mu $ e 
$K^\mu =\frac 12(k_1^\mu +k_2^\mu )$.

In the case of freeze out (in the Bjorken model with 
pseudo-temperature\cite{ko86})
\begin{equation}
G(k_1,k_2)=2<\frac{dN}{dy}> \{\frac 2{q_TR_T}J_1(q_TR_T)\}K_0(\xi ) 
\end{equation}
where
\begin{eqnarray}
\xi^2&=&[\frac 1{2T}(m_{1T}+m_{2T})-i\tau(m_{1T}-m_{2T})]^2+ \nonumber\\
&&2\,(\frac 1{4T^2}+\tau ^2)\,m_{1T}\,m_{2T}\,[\cosh (\Delta y)-1]\;\;,
\label{xi}
\end{eqnarray}
$\Delta y=y_1-y_2$, $<\ >$ indicates average over particles 1 and 2\\ 
and 
\begin{equation}
G(k_i,k_i)=2\frac{dN}{dy_i}K_0(\frac{m_{iT}}T).
\end{equation}

In the case of continuous emission
in the   Bjorken model  with 
pseudo-temperature
\begin{eqnarray}
G(q,K)& = &\frac 1{(2\pi )^3(1-{\bf {\cal P}}_{{\cal F}})}\int_0^{2\pi }d\phi
\int_{-\infty }^{+\infty }d\eta  \nonumber \\
& \times & \{\int_0^{R_T}\rho \;d\rho \;\tau _{{\cal F}}\;M_T\;\cosh (Y-\eta)
\nonumber \\
&\times &  e^{i[\tau _{{\cal F}}(q_0\cosh \eta -q_L\sinh \eta )-\rho q_T\cos
(\phi -\phi _q)]}  \nonumber \\
& + & \!\int_{\tau _0}^{+\infty }\!\!\tau d\tau \rho _{{\cal F}}K_T\cos \phi 
\nonumber \\
&\times & \,e^{i[\tau (q_0\cosh \eta -q_L\sinh \eta )-\rho _{{\cal F}}q_T\cos
(\phi -\phi _q)]}\}  
\nonumber \\
& \times &  e^{-M_T\cosh (Y-\eta )/T_{ps}(x)}, \nonumber \\
\end{eqnarray}
with
$M_T=\!\sqrt{K_T^2+M^2},\;\vec{K}_T=\frac 12 
(\vec{k}_1+\vec{k}_2)_T\,,\,M^2=K_\mu K^\mu=m^2-\frac 14q_\mu q^\mu $,
$Y$ is the rapidity corresponding to  $\vec{K}$, $\phi $ is
 the azimuthal angle in  
 relation to the  direction of   $\vec{K}$ and  $\phi _q$ is the 
 angle between the directions of  $\vec{q}$ and $\vec{K}$. $\tau_{\cal{P_F}}$ and
$\rho_{\cal{P_F}}$ are determined by $\cal{P}=\cal{P_F}$. 
$T_{ps}(x)=1,42 T(x)-12,7 MeV$.

In \cite{Grassi00b}, a few idealized cases were studied and then 
some cases more representative of the experimental situation, were presented. For example,
instead of $C(q,K)$, we computed
\begin{eqnarray}
 & \langle C(q_L)\rangle  = 1+  \nonumber \\ 
 & \frac{\int_{-180}^{180}dK_L\int_{50}^{600}dK_T\int_0^{30}dq_S
  \int_0^{30}dq_oC(K,q)|G(K,q)|^2}{\int_{-180}^{180}dK_L
  \!\int_{50}^{600}dK_T\!\int_0^{30}dq_S\!\int_0^{30}dq_o 
  C(K,q)G(k_1,k_1)G(k_2,k_2)}.  \nonumber \\
 &  
\end{eqnarray}
(This corresponds to the experimental cuts of  NA35).
$q_O$, $q_S$ e $q_L$ are defined in figure~\ref{fig:def}.
\begin{figure}[htbp]
\begin{center}
\epsfig{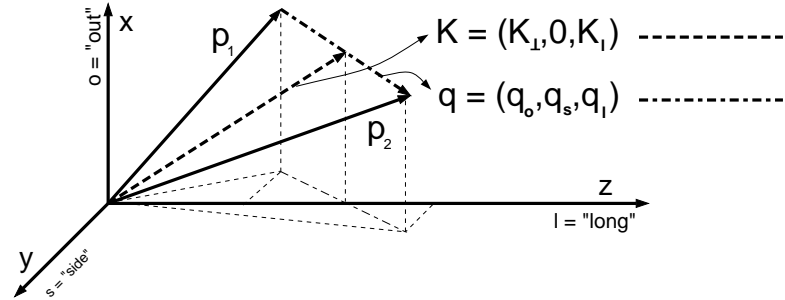}
\end{center}
\caption{ By definition : $Oz$ is chosen along the beam and
$\vec{K}$ in the $x-z$ plane.  The  $L$ (longitudinal) component of a vector is
its 
 $z$ component,  the $O$ (``outward'') its $x$ component and $S$
(``sidewards'') its  $y$ component \cite{dessinhbt}.
\label{fig:def}}
\end{figure}

In a first comparison, we used similar initial conditions than above,
 $T_0=200$ MeV  (S+S collisions) for both continuous emission and freeze out.
 Results are presented as function of
 $q_L$,$q_O$ and $q_S$ in figure~\ref{fig:HBT1}.
\begin{figure}[htbp]
\begin{center}
\epsfig{file=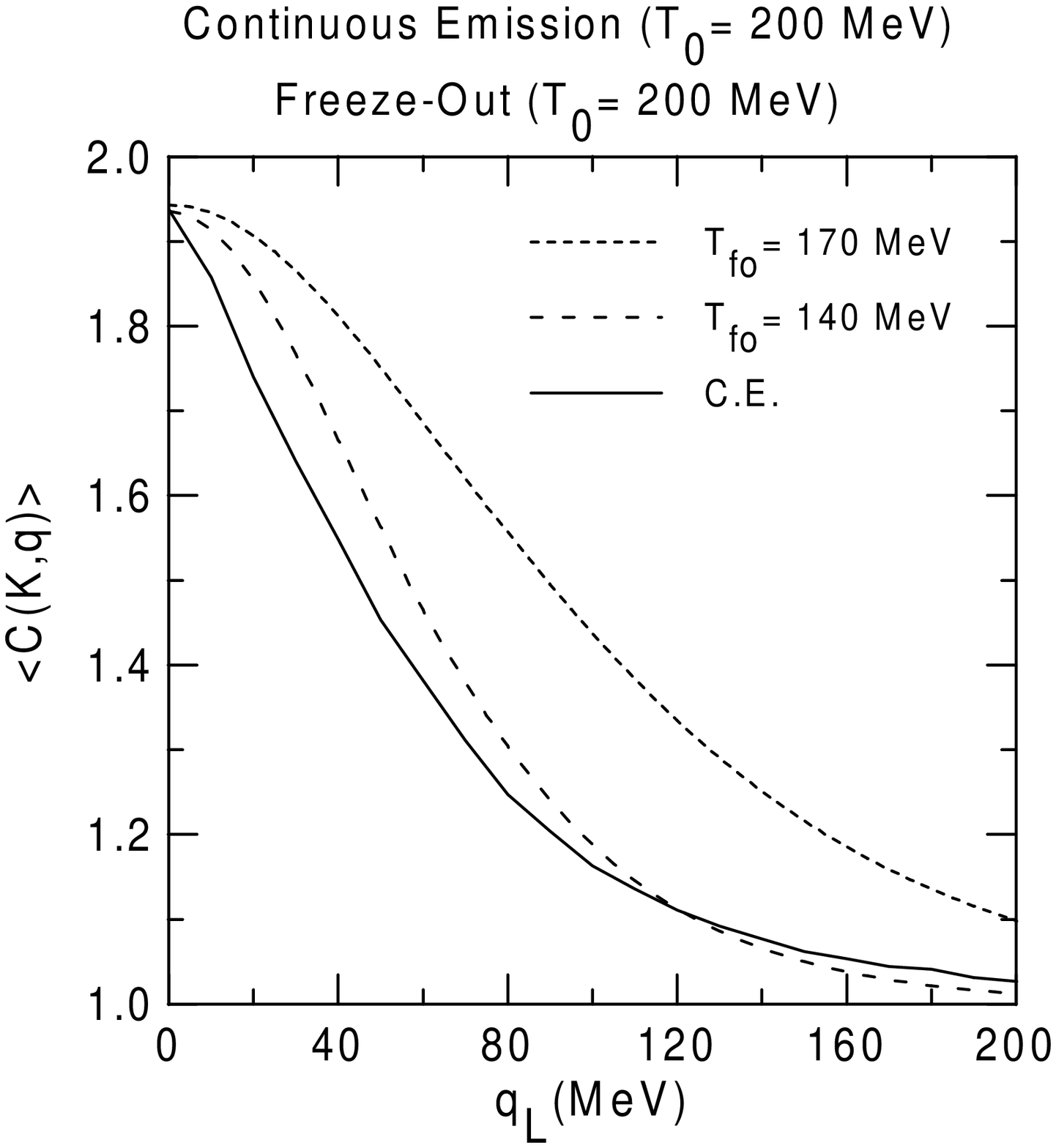,height=7.cm}\hfill\epsfig{file=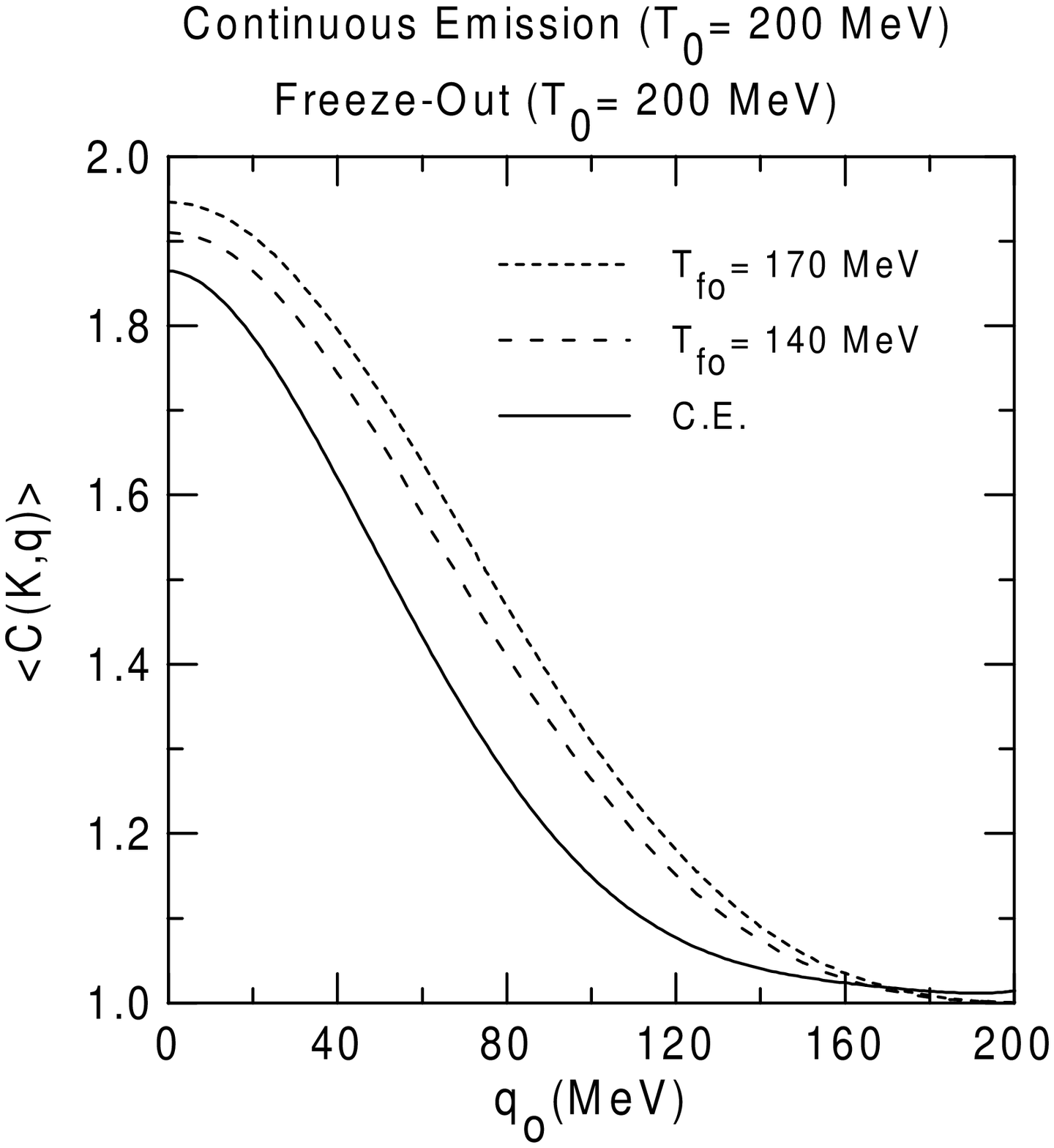,height=7.cm}
\hfill\epsfig{file=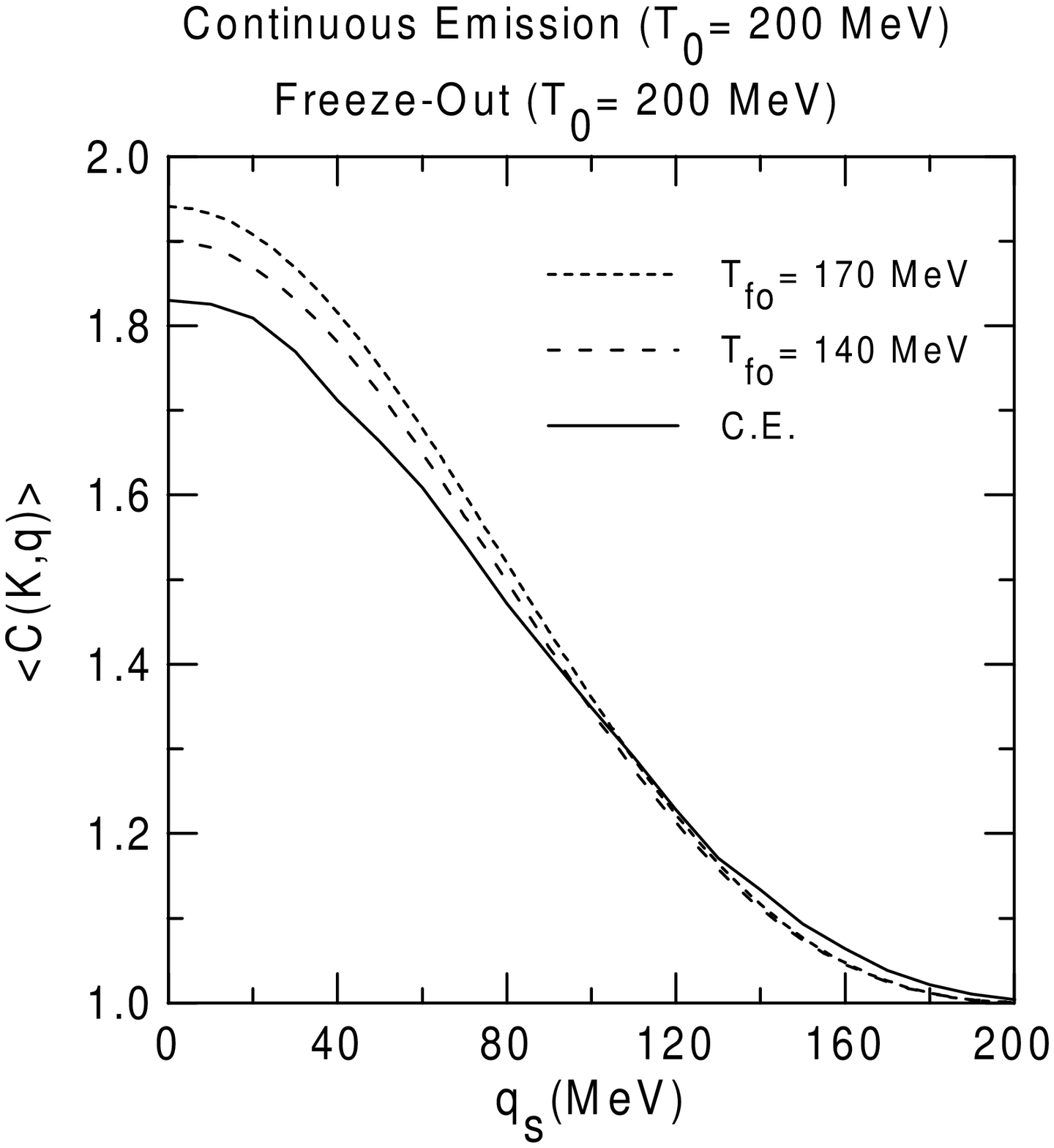,height=7.cm}
\end{center}
\caption{ Comparison of continuous emission with  freeze out, 
when both have  the same initial conditions  \cite{Grassi00b}.
\label{fig:HBT1}}
\end{figure}
In a second comparison, given a curve obtained for continuous emission,
we try to find a similar curve obtained with the standard 
value
$T_{f.out}=140$ MeV, varying the initial   temperature $T^{f.out}_0$.
The results are shown as function of
$q_L$,$q_O$ e $q_S$ in figure~\ref{fig:HBT2}.
\begin{figure}[htbp]
\begin{center}
\epsfig{file=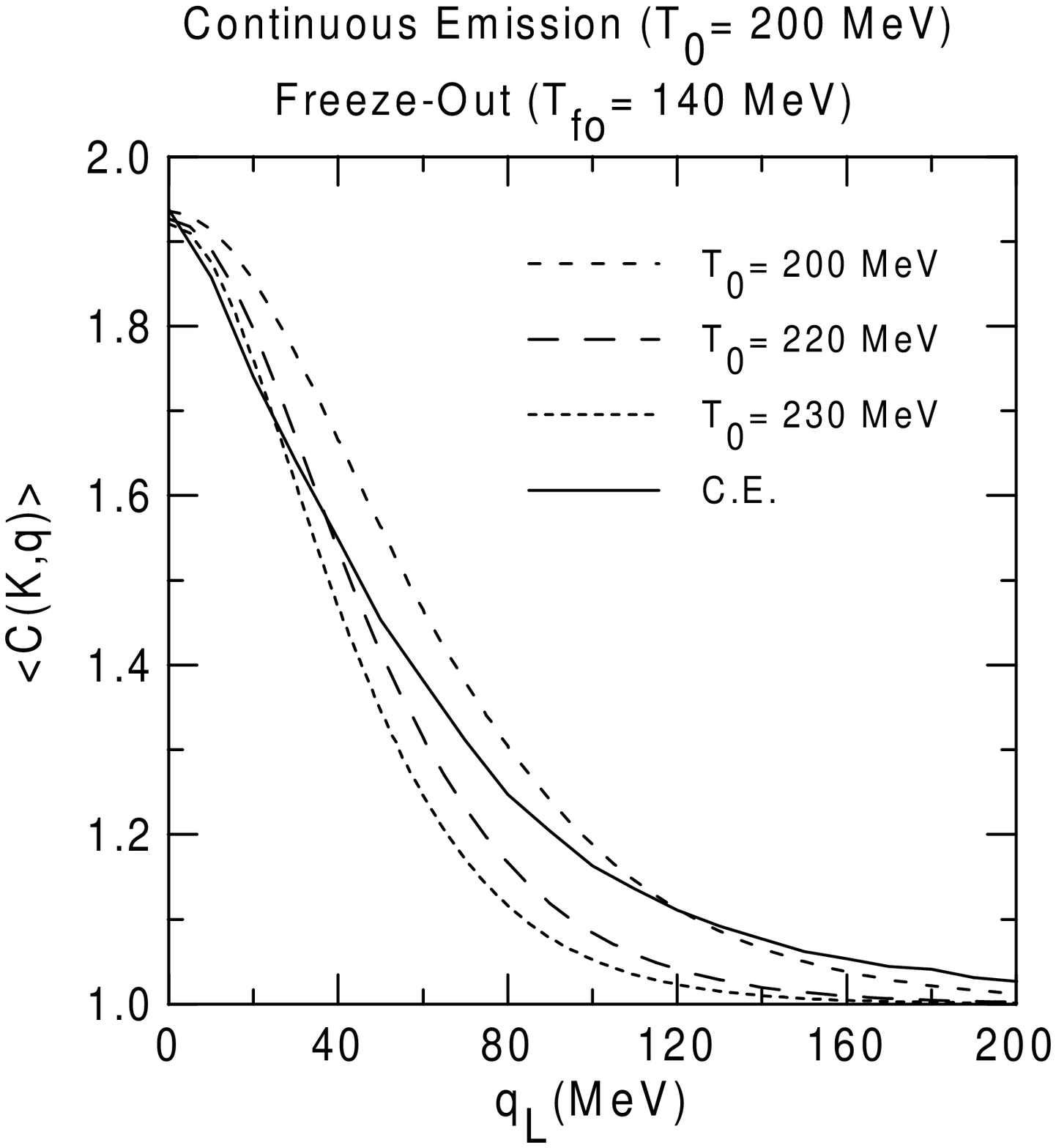,height=7.cm}\hfill\epsfig{file=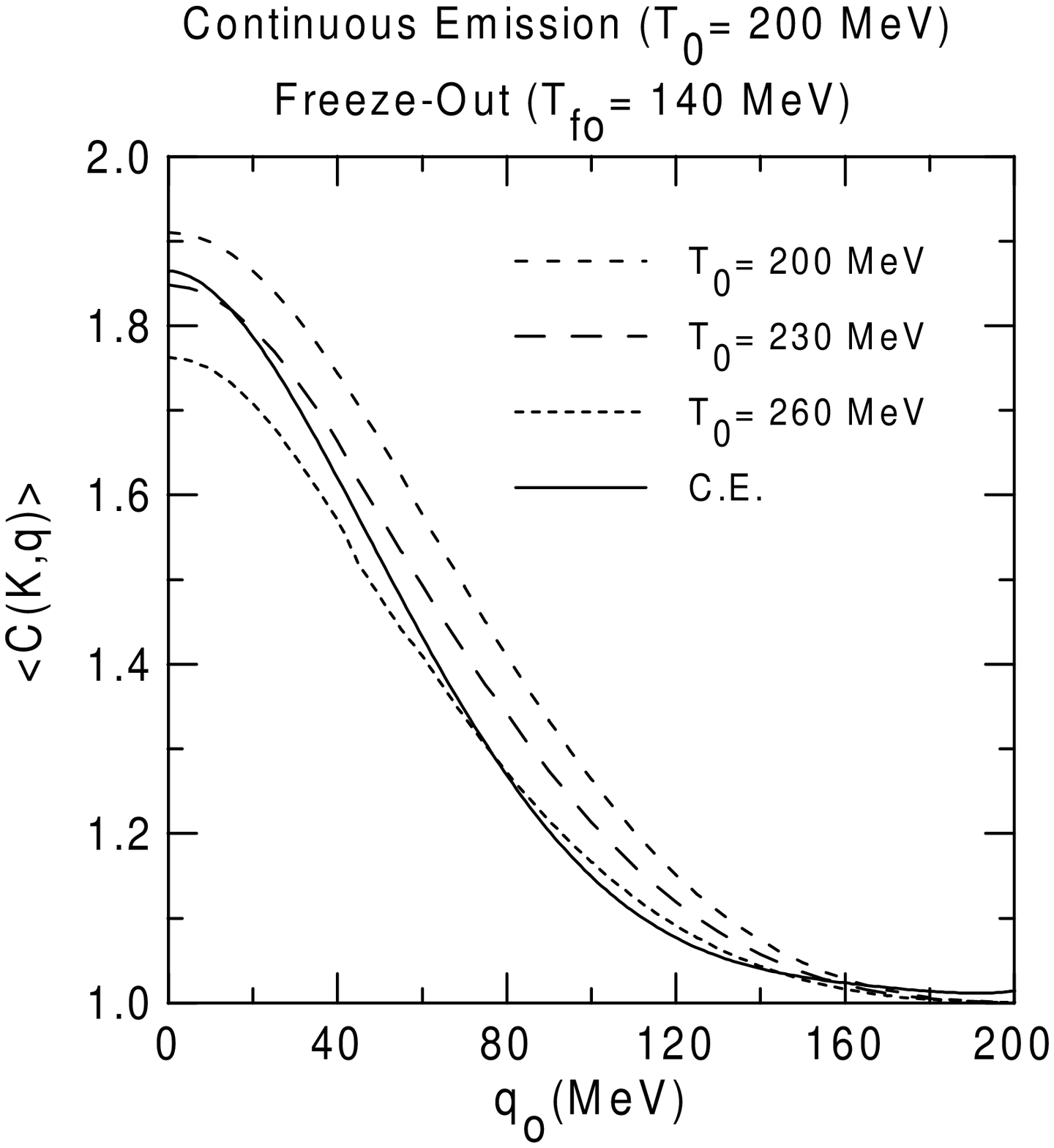,height=7.cm}
\hfill\epsfig{file=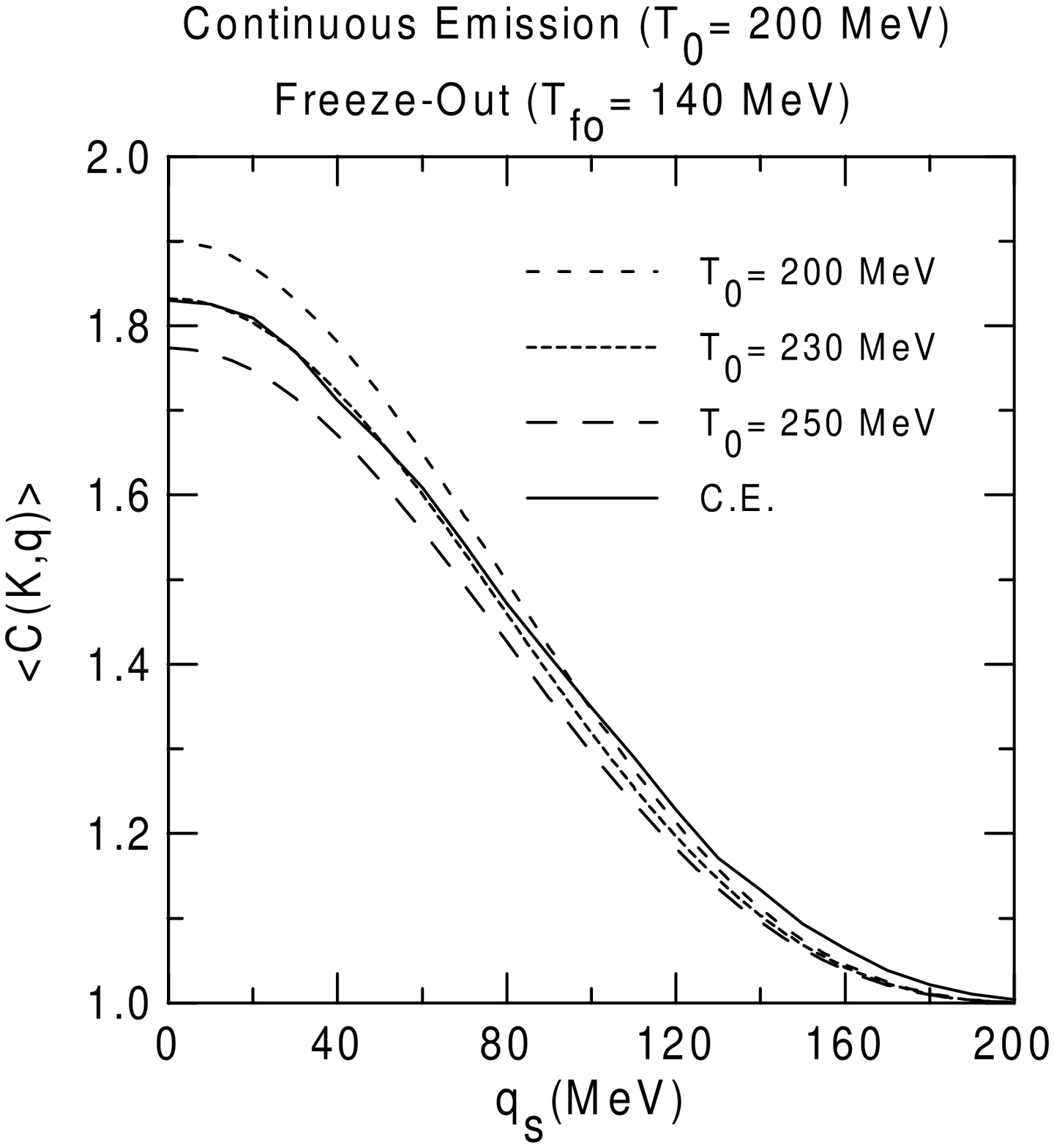,height=7.cm}
\end{center}
\caption{ Given a curve obtained for continuous  emission, 
we look for initial conditions for freeze out at $T_{f.out}=140$ MeV leading to a similar curve
\cite{Grassi00b}.
\label{fig:HBT2}}
\end{figure}
From these two sets of figures,
it can be seen that there are many differences in the
correlations between both
models: shape. heigth,  etc. If the initial conditions are the same,
 the correlations are very different. Even more interesting
if trying to approximate with a freeze out at $T_{f.out}=140$ MeV,
the continuous emission  correlations, it is necessary to assume a  $T_0$ very high, where the notion of  hadronic gas looses its validity.
So, we can see that if continuous emission is the correct description for
 experimental data, it will be more difficult to attain the quark gluon plasma
than it looks  using the freeze out   model.
So again we conclude that what we learn about the hot dense matter created depends on the emission model used.
( Note that to actually compare with data, transverse expansion  has to be included. )

 More recently, we addressed the ``HBT puzzle'' at RHIC using NeXSPheRIO~\cite{hbt04} (for a review on NeXSPheRIO, see \cite{yogiro_review}).
We showed that the use of varying initial conditions leads to smaller radii.
In addition, though continuous emission is not easy to introduce in hydrodynamical codes (because (\ref{eq:glauber}) depends on the future evolution of the fluid), 
it was introduced in an approximate way and shown to be important
to reproduce the momentum dependence of the radii.
\\

{\bf f. Plasma}\\
In all the previous analysis, we assumed that the fluid was initially composed of hadrons with some initial conditions
$T_0$, $\mu_{b,0}$.  
Given the possibility that a quark gluon plasma might have been created already, we must discuss the extension of our model to the case were a plasma might have been formed.

Initially, it might be expected that continuous emission by a plasma is
impeded for two reasons:
 1) a hadron emitted by the plasma surface, in the outward direction,  
must cross all the hadronic matter around the plasma core , so   probably it will suffer collisions and when finally emitted, it will be emitted by the hadronic gas in the way we have considered so far
 2) the plasma is
formed by colored objects that must recombine in a color singlet
at the plasma surface  to be emitted, this makes plasma emission more difficult than hadron gas emission.

For simplicity let us consider a second order phase 
transition.
We use for the hadron gas a  resonance gas  equation of state. For the plasma, we use a MIT bag equation of state where the value of the bag constant
and transition temperature are  adjusted
to get a second order phase transtion.
We get $T_c \sim 220$ MeV and $B \sim  580$ MeV $fm^{-3}$.

Then we solve the hydrodynamics equation (without continuous emission as a first approximation)
to know the localization and evolution of the plasma core.
This is shown in figure~\ref{fig:qgp}. 
These equations were also solved for the case of a hadronic gas for comparison in figure~\ref{fig:hg}.
\begin{figure}[htbp]
\epsfig{file=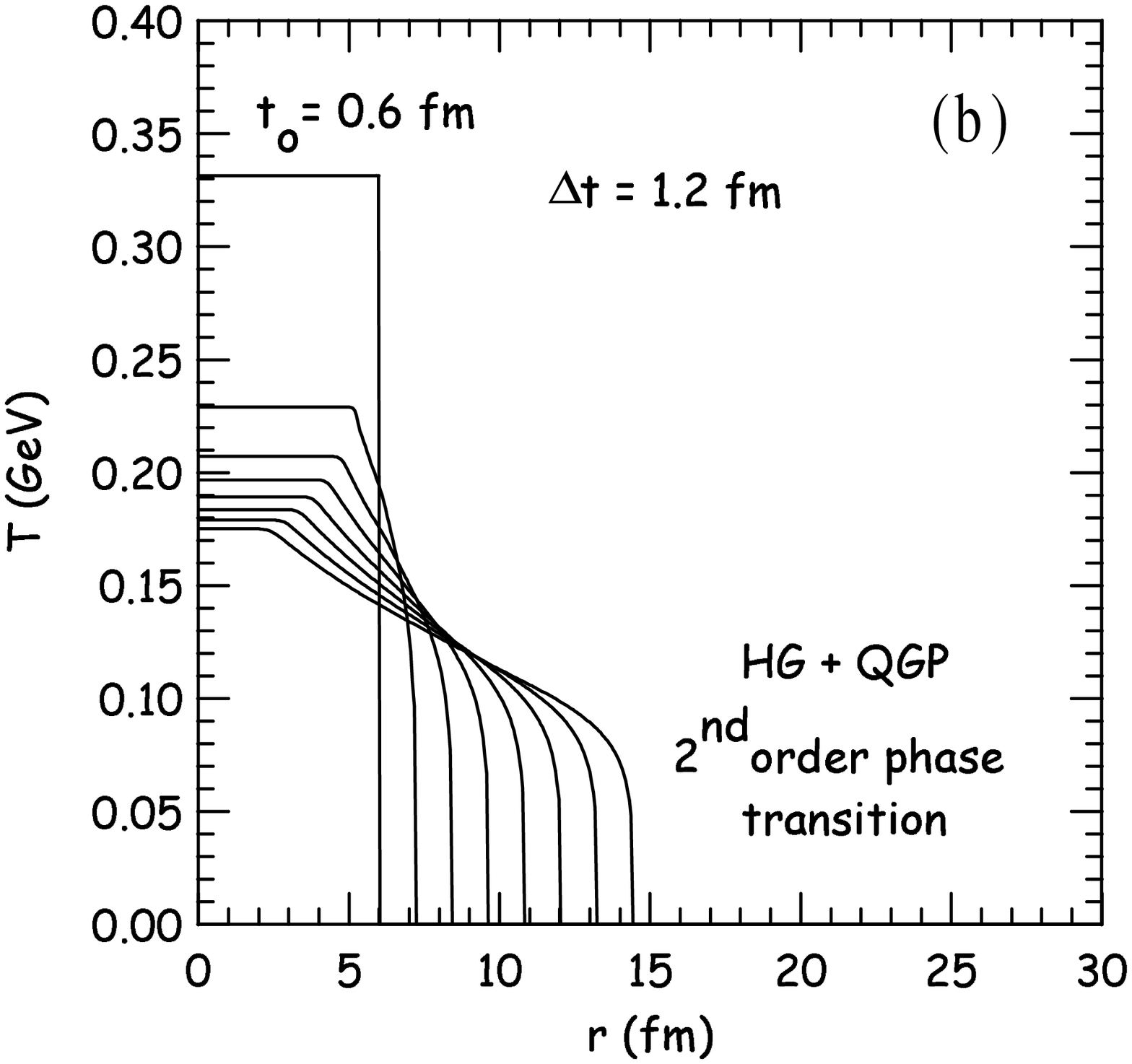,height=6.cm,angle=0}
\epsfig{file=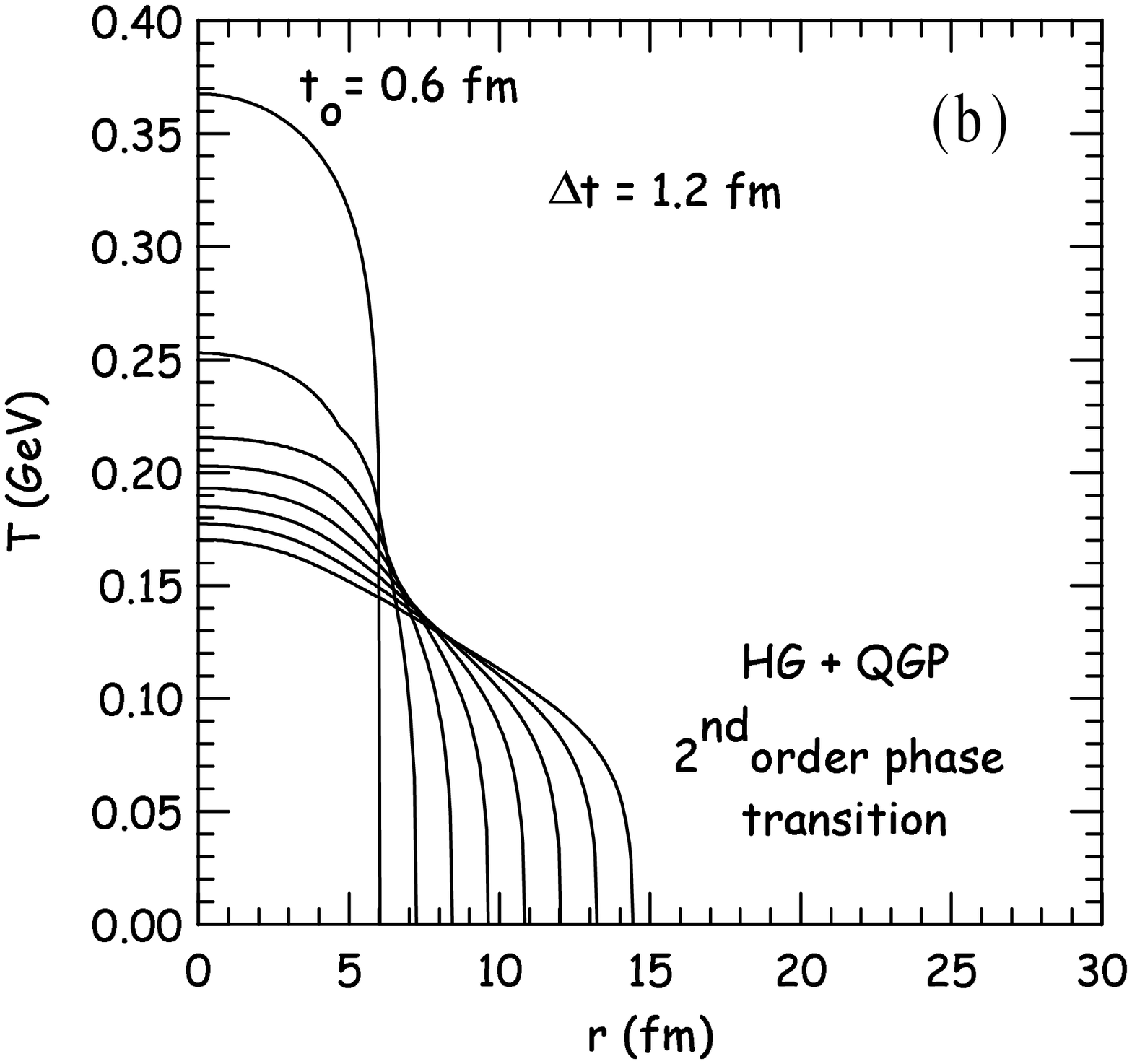,height=6.cm,angle=0}
\caption{ Evolution of temperature in  the case
of a hadron gas with quark core
 and initial boxlike
 matter distribution,  top and softer, bottom.
\label{fig:qgp}}
\end{figure}
\begin{figure}[htbp]
\epsfig{file=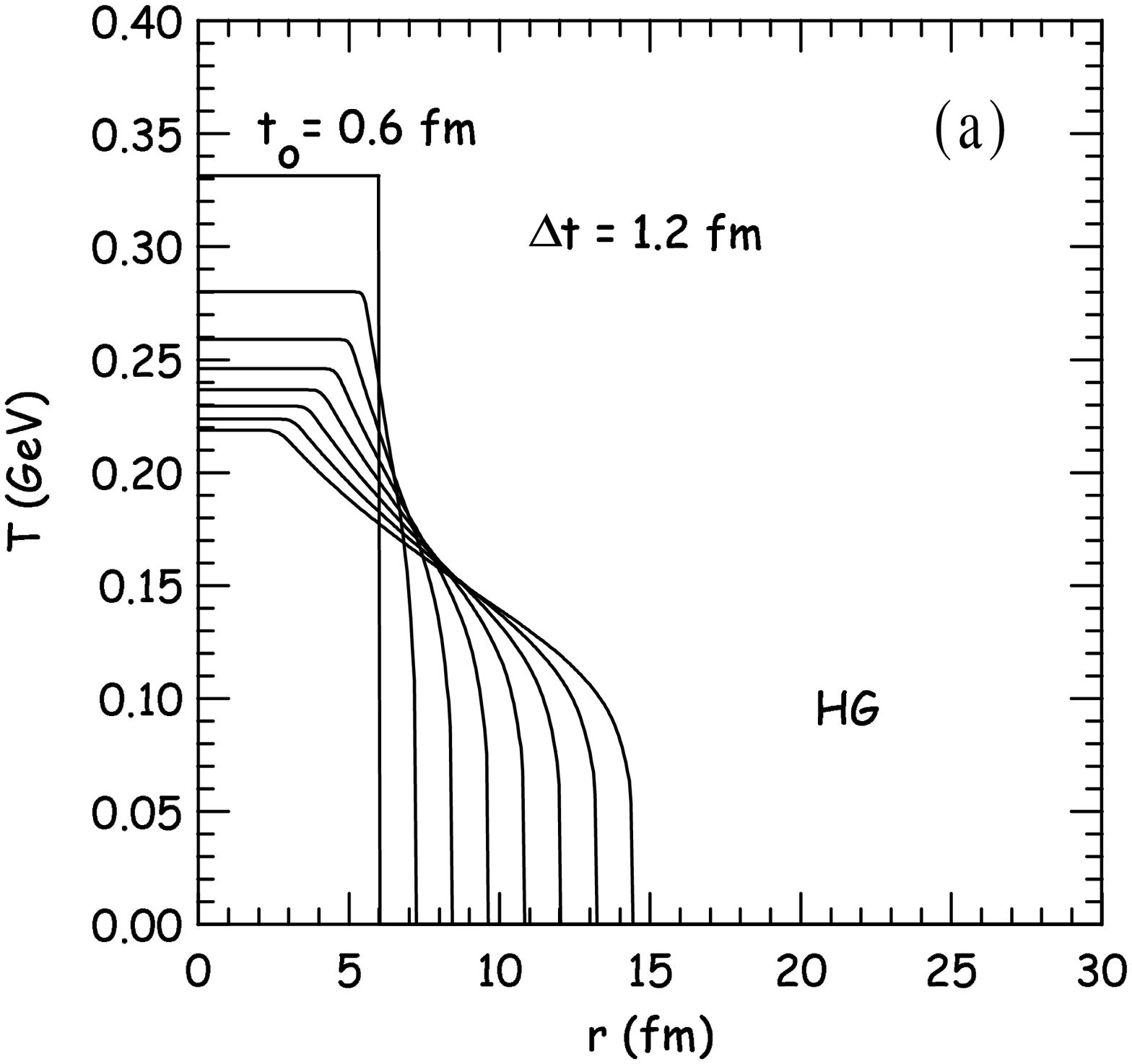,height=6.cm,angle=0}
\epsfig{file=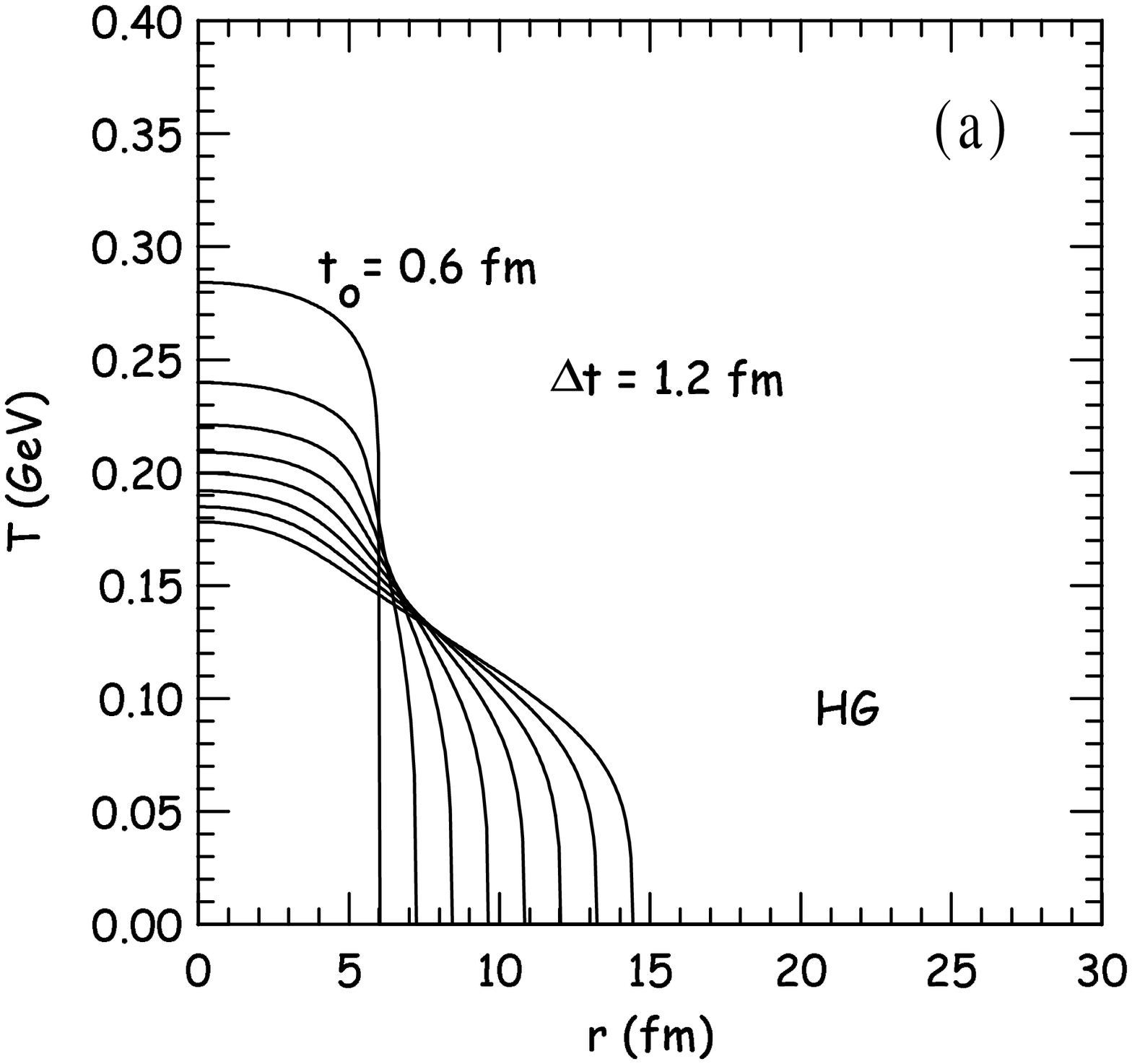,height=6.cm,angle=0}
\caption{ Evolution of temperature in  the case
of a hadron gas 
 and initial boxlike
 matter distribution,  top and softer, bottom.}
\label{fig:hg}
\end{figure}
It can be checked that when there exists a plasma core, it is quite close to the outside region. Contrarely to the reservation 1) above,
hadrons emitted by the plasma surface might be quite close enough 
to the outside to escape without collisions.

Now for reservation number 2), we note that there exist various mechanisms
\cite{raf83,ban83,mu85,vi91,russe} proposed for hadron emission by a plasma.
To start we can assume as  Visher et al. \cite{vi91} that the plasma emits in equilibrium with the  hadron gas.
In this case, the emission formula by the plasma core+hadron gas would be
\begin{eqnarray}
E\frac{d^3N}{dp^3} & = & \int d\phi d\eta 
[ \int_0^{\infty} \rho d\rho
\left( p^{\tau} f_{free} \tau \right)_{|\tau_{\infty}}  \nonumber \\
& + &
\int_{\tau_0}^{\infty} \tau d\tau
\left( p^{\rho} f_{free} \rho \right)_{|R_{out}(\tau)} ]
\end{eqnarray}
where $R_{out}(\tau)$ is the radius up to which there is matter. \\
This is similar to the hadron gas case treated above. The difference is in the calculation of
 $\cal P$, (which appears in  $f_{free}$),
since a hadron entering the plasma core will be supposed detroyed.

In the same spirit as above,  a cutoff at 
${\cal P}={\cal P_F}$ can be introduced.
Due to the similarity for the spectra formula with and without plasma core,
we do not expect very drastic differences if the transition is second order.
Of course, the case of first order transition must be considered (though results from lattice QCD on the lattice do not favour strong first order transition).
(Note that $T_c \sim 220$ MeV is higher than expected, this might be improved
e.g. using a better equation of state).

\section{Conclusion}

In this paper,  we discussed particle emission in hydrodynamics.
Sudden freeze out is the mechanism commonly used.
We described some of its caveats and ways out. 

First the problem of negative contributions in the Cooper-Frye formula was presented.
When these contributions are neglected, they lead to violations of conservation laws. We showed how to avoid this, the main difficulty remains to compute the distribution function of matter that crossed the freeze out surface.
Even models  combining hydrodynamics with a cascade code have this type of problem or related ones~\cite{bu03}.

Assuming that sudden freeze out does hold true, data call for two separate freeze outs. 
We argue that in this case, this must be included in the hydrodynamical code
as it will influence the fluid evolution and the observables.
Some works \cite{flor} using paramatrization of the hydrodynamical solution 
suggest that a single freeze out might be enough.
No  hydrodynamical code with simultaneous chemical and thermal freeze outs
achieves this so far (see e.g. our figures 
~\ref{fig:hylander}). 
On the other side, (single) explosive freeze out is being incorporated in a hydrodynamical code \cite{csernaicode}.

Finally, we argued that microscopical models indeed do not indicate a sudden freeze out but a continuous emission~\cite{Bra,Sor,Bas}. 
We showed how to formulate particle emission in hydrodynamics for this case and
discussed extensively its consistency with data. We pointed out in various cases
 that the interpretation of data is quite influenced by the choice of the particle emission mechanisms.
The formalism that we presented for continuous emission needs improvements, for example the ansatz of immediate rethermalization is not realistic.
An example of such an attempt is~\cite{yogiro}.
Finally, it is also necessary to think of ways to include continuous emission 
in hydrodynamics. This is not trivial because the probability to escape depends on the future. In \cite{hbt04}, such an idea was applied to the ``HBT puzzle'' at RHIC with promising results.

\subsection{Acknowledgments}
This work was partially supported by FAPESP (2000/04422-7).
The author wishes to thank L.Csernai and S.Padula for reading parts of the 
manuscript prior to submission.

\end{document}